\documentstyle[12pt,epsfig,amsmath,bezier,amssymb]{article}
\begin{document}

\title{\bf Glassy systems under time-dependent driving forces: \\
application to slow granular rheology}

\author{L. Berthier$^1$, L. F. Cugliandolo$^2$
 and J. L. Iguain$^{3,\star}$
\\
 $^1${\it Laboratoire de Physique, \'Ecole Normale Sup\'erieure de 
 Lyon,}  
\\
{\it 46 All\'ee d'Italie,  F-69007 Lyon, France and}
\\
{\it D\'epartement de Physique des Mat\'eriaux, U. C. B. Lyon 1,}
\\
{\it 69622 Villeurbanne Cedex, France}
\\
$^2${\it Laboratoire de Physique Th\'eorique, \'Ecole Normale 
Sup\'erieure,} 
\\
{\it 24 rue Lhomond, F-75231 Paris C\'edex 05, France and }
\\
{\it Laboratoire de Physique Th\'eorique et Hautes \'Energies, Jussieu,}
 \\
{\it 4 Place Jussieu,  F-75252 Paris C\'edex 05, France}
\\
$^3${\it Departamento de F\'{\i}sica, 
Universidad Nacional de Mar del Plata,} 
\\
{\it De\'an Funes 3350, 7600 Mar del Plata, Argentina}
}

\date{\today}

\maketitle

\begin{abstract}
We study the dynamics of a glassy model with infinite 
range interactions externally  driven by an 
oscillatory force.  We find a well-defined transition in the 
(Temperature-Amplitude-Frequency) phase diagram 
between (i) a `glassy' state
characterized by the slow relaxation of one-time quantities, 
aging in two-time quantities and a modification 
of the equilibrium fluctuation-dissipation relation; 
and (ii) a `liquid'
state with a finite relaxation time.
In the glassy phase, the degrees of freedom governing the 
slow relaxation
are thermalized to an effective temperature.
Using Monte-Carlo simulations, we investigate the effect 
of trapping regions in phase space on the driven dynamics. 
We find that it alternates between periods of rapid motion and periods 
of trapping.
These results confirm the strong analogies between the slow granular 
rheology and the dynamics of glasses. They 
also provide a theoretical underpinning to 
earlier attempts to present a thermodynamic description
of moderately driven granular materials.
\end{abstract}
 
\vspace{.25cm}

\noindent
$^\star$Present address: LPTHE, 4 Place Jussieu, 
F-75252 Paris Cedex 05, France.

\noindent
LPTENS 00/34, LPTHE 00/36, LPENSL-TH-13/2000.

\noindent 

\newpage

\section{Introduction}
\label{introduction}

In recent years, granular matter has received a growing attention
from the physics community \cite{reviews}.
The study of powders is relevant not only because granular materials
have many industrial applications but also because it raises 
many fundamental questions of physical interest.

We call granular matter all many-body systems constituted by
grains of macroscopic size. 
The grains interact via repulsive dissipative forces due to inelastic
collisions and static friction. The thermal energy scale is totally
negligible with respect to the typical gravitational energy.
In the absence of external perturbations, each metastable 
configuration has an infinite life-time and thermal averaging
is meaningless.
The static properties of such systems are hence
very interesting, the sandpile problem being the paradigm \cite{reviews}.

Powders flow only when  energy is supplied externally. This can be 
done by applying a shear or a vibration.
The dynamics of granular matter presents 
a very rich  phenomenology that 
depends not only on the intensity of the drive,
generically called $\Gamma$, 
but also on the 
way the granular system is driven~\cite{reviews}. 
A weak driving force can be provided  by applying
`taps' to the systems, as has been done in the pioneering experiments of 
the Chicago group \cite{compaction_exp}, where the parameter 
$\Gamma$ is the reduced acceleration of the taps.
Recently, Nicolas {\it et al.} investigated the dynamics 
of a powder by sinusoidally
shearing it in a weak manner~\cite{pouliquen}; 
$\Gamma$ is here the 
maximal amplitude of the strain. 
Experiments reveal that the time evolution, 
in the gently driven situation, is
characterized by an extremely slow dynamics
\cite{compaction_exp,pouliquen,jaeger}. 
In the tapping experiments for instance, the density
still evolves after $10^5$ taps \cite{compaction_exp}.
When the energy injection is much stronger, the granular matter 
eventually  becomes fluid, it behaves essentially
like a dissipative gas and it is described
by a hydrodynamic theory that takes into account
energy dissipation through the collisions between 
the grains~\cite{reviews,fauve}.
In this work we shall focus on the gently driven regime and we 
shall not address the strongly perturbed situation.

This phenomenology is clearly reminiscent of the behavior of 
glass forming systems for which the control parameter 
is the temperature $T$
(or the density $\rho$) \cite{edwards1,liu_nagel,mehta,mehta-barker,jorge}.
At high $T$ (say), the samples are in the liquid or gaseous phases.
When $T$ decreases, their dynamics becomes exceedingly slow,
and may even appear completely stopped during the experimental 
time-window. 
However, at temperatures above but close to the glass transition
the relaxation reaches a stationary regime characterized by the decay
of all correlations in two steps, the second decay being related to the  
structural relaxation.
At temperatures below the glass
transition temperature, the structural relaxation time $t_r$ 
depends on the time spent 
in the glassy phase (the `waiting time' $t_w$, or `age') and  typically
$t_r \propto t_w$~\cite{review_aging}.
A stationary regime cannot be reached experimentally.
This is the aging effect which has been observed in 
a wide spectrum of glassy systems such as plastics~\cite{Struick},
spin-glasses \cite{spin-glasses}, 
glycerol~\cite{Leheny}, dielectric glasses~\cite{Levelut}, 
complex fluids~\cite{gel}, 
phase separating systems~\cite{Bray}, etc. 

Recently, the similitude in the dynamics of granular matter under 
vibration and glass forming materials has been rationalized by 
Liu and Nagel \cite{liu_nagel}.
These authors proposed a phase diagram that unifies
the physics of glassy systems and granular materials.
In its simplest version, the diagram has three axis 
($T$, $\rho$, $\Gamma$). 
The ($T$, $\rho$) plane describes the physics of glasses, while
the ($\rho$, $\Gamma$) plane describes the one of athermal driven systems,
like powders or foams.
In the low-$T$, high-$\rho$, small-$\Gamma$ region
the system is generically jammed or presents glassy features.
In this work, we focus on 
the $(\Gamma,T)$ plane of this phase diagram.
The drive axis $\Gamma$ 
can represent two types of forces:
(1) `shear-like' forces that  
do not derive from a potential and hence do work on the 
sample \cite{Cukulepe};
(2) `tapping-like' forces that 
do derive from a potential but do work on the samples when they 
depend on time.
Both modify the dynamic behavior and the goal is to identify
how, and to which extent, in a general manner.
One possible scenario is that 
the age of the driven system stabilizes
at a power dependent level, typically $t_r \propto \Gamma^{-1}$.
In the rheological language, this is a shear-thinning behavior.
Some examples 
are given by domain growth under flow~\cite{Onuki} 
or by the non-linear rheology of complex fluids
\cite{petrol}, and it has been captured
by a number of models \cite{BBK,peter,caroline}.
In particular, a (Shear/Temperature) phase diagram for glassy
systems has been derived in Ref. \cite{BBK}.
Another scenario, realized in the present paper, is 
that aging is not stopped, at least 
in a  well-delimitated region 
of the phase diagram.

The behavior of  moderately driven granular matter has received a lot
of experimental \cite{compaction_exp,pouliquen,jaeger,josserand,anna}, 
numerical \cite{mehta-barker,tetris0,tetris1,alain,mauro} and theoretical 
\cite{mehta,jorge,boutreux,parking,linz,hong,gavrilov,head1,head2,edwards2,edwards} attention. 
All these studies have demonstrated the glassy 
nature of granular compaction below a critical amplitude of 
the drive, $\Gamma^\star$. This has been  first revealed 
by the very slow relaxation of the density, but 
memory experiments \cite{josserand} and simulations 
\cite{alain,parking} inspired by earlier spin glass 
studies \cite{review_aging,spin-glasses} have also
given support to this conclusion.

The relation between granular matter and glassy systems 
is widely assumed. Indeed, many  of the models which have been proposed to
study granular compaction are directly
adapted from 
glassy models~\cite{mehta,tetris0,tetris1,alain,mauro,head2,edwards2,edwards}.
Usually, the drive $\Gamma$ in granular matter is related to the 
temperature $T$ in glasses.
However, the assumption that $T = T(\Gamma)$ is a highly 
non-trivial statement and
there is, to our knowledge, no microscopic approach 
that justifies it. {\it We do not make such an assumption here.}
In this respect,  Mehta {\it et al.} \cite{mehta} have 
built a phenomenological two-temperature 
stochastic model based on the observation that the slow granular compaction
is basically a two-step process: in this model, a short-time process 
stands for the fast independent-particle relaxation, while 
a slow one stands for cooperative rearrangements.
The recent experiment of Nicolas {\it et al.} \cite{pouliquen}
clearly proved the existence
of these two (uncorrelated) processes.
A two-step process, each thermalized with its own temperature 
is precisely the output of previous studies of the
constantly driven dynamics of glassy systems in a thermal 
bath~\cite{Cukulepe,BBK}, confirmed
by the numerical simulation of a sheared supercooled liquid \cite{BB}.

In this paper we study the dynamics of a glassy system
permanently perturbed by a time-dependent Hamiltonian
force~\cite{comment1}. 
Our aim is to identify which properties correspond to those observed 
experimentally in granular systems and whether an effective temperature
for the slow degrees of freedom is generated in
this weakly athermal system.
In some sense, our approach is `orthogonal' to previous ones.
We do not propose a new model for the slow granular rheology, 
but rather ask the following simpler question: {\it What is 
the behavior of a glassy system subjected to periodic driving forces?}
To answer this question, the ($T$, $\Gamma$, $\omega$)
phase diagram is explored. 
This is done by studying the dynamics of the, by now standard,
mean-field glass model, the $p$-spin glass 
model~\cite{review_aging,Kith,Crso},  
under a time-dependent driving force. [Albeit its name, that follows from
historical reasons, this model represents a structural glass and not 
a spin glass.]
We then discuss how our results have to
be interpreted in the context of granular materials.

The paper is organized as follows. In the next Section, we
recall the definition and main properties 
of the $p$-spin glass model.
Its behavior under an oscillatory drive, for the choice of parameters 
$p=2$ and $p \ge 3$, is examined in Sections~\ref{drivenp=2}
 and \ref{drivenp=3}, respectively, 
in two ways: analytical and numerical solutions of the 
spherical model, and Monte-Carlo simulations of the Ising model.
We discuss our results in the context of slow 
granular rheology in Section~\ref{discussion}.
Our conclusions close the paper in Section~\ref{conclusion}.

\section{Definition and zero-drive behavior of the model}
\label{definition}

The spherical $p$-spin glass, when $p=2$, 
is simply the spherical version of the 
Sherrington-Kirkpatrick spin glass and 
it was introduced by Kosterlitz {\it et al.} \cite{KTJ} 
as an exactly solvable model.
It is in fact equivalent to the
$O(n)$, $n \rightarrow \infty$, model for 
ferromagnetic domain growth in three dimensions.
Its statics \cite{KTJ} and  its relaxational dynamics have been extensively
studied~\cite{p=2,cude,cude2}.
The $p \ge 3$ spherical models instead are simplified models for supercooled
liquids and glasses, in the sense that they give a theoretical framework
to understand both the statics \cite{review_aging,Kith,mepa} and the 
non-equilibrium dynamics~\cite{review_aging,Cuku} of glass forming systems.

We study these two cases as generic glassy models, taking advantage 
of the fact that exact equations
for their driven dynamics can be derived.

\subsection{Model}

The $p$-spin glass model in its spherical \cite{Crso} and 
Ising \cite{Kith,De} versions is  defined by the Hamiltonian
\begin{equation}
 H_J[{\bf s}] =  
\sum_{i_1 < i_2 < \dots <i_p } J_{i_1 i_2\dots i_p} s_{i_1} s_{i_2}
 \dots s_{i_p} \;,
\label{ham_pspin}
\end{equation}
where the couplings 
$J_{i_1i_2\dots i_p}$ are random Gaussian variables with zero mean and
 variance $\overline{({J_{i_1 i_2\dots i_p}})^2} = p!/(2N^{p-1})$.
The spins $\{s_i,i=1,\cdots,N\}$ may satisfy a global 
spherical constraint $\sum_i {s_i}^2(t)=N$, or be Ising 
variables $s_i=\pm 1$.
In the spherical version, the driven dynamics is given by the following
Langevin equation
\begin{equation}
\frac{\partial s_i (t)}{\partial t} = -\frac{\delta H}{\delta s_i(t)} 
-z(t)s_i(t) + f_i^{\sc{tapping}}(t) + \eta_i(t) \; ,
\label{langevin}
\end{equation}
where the parameter $z(t)$ ensures the spherical constraint and 
$\eta_i(t)$ is the thermal noise, taken from a Gaussian distribution
with zero mean
and variance $2k_{\rm B}T$.
[In what follows we use units such that the Boltzmann 
constant $k_{\rm B}$ is one.]
The thermal bath temperature, $T$, can
possibly be zero.
In the Ising version one can still write down a continuous Langevin
equation by using soft spins and taking the Ising limit at the end of the 
calculations.  
The spherical version is simpler to treat analytically
while the Ising version is much simpler to deal with 
using Monte-Carlo simulations, since the spins are bimodal variables. 

In order to mimic tapping experiments, 
a periodic time-dependent Hamiltonian force  $f_i^{\sc{tapping}}(t)$
has to be added to the right hand side of the Langevin 
equation (\ref{langevin}).
The simplest periodic time-dependence one can think of is a cosine form 
of period $\tau \equiv 2\pi/\omega$.
Thus, we are lead to add a magnetic field in the Hamiltonian (\ref{ham_pspin}),
\begin{equation}
H_J[{\bf s}] \to H_J[{\bf s}] + \cos(\omega t)
 \sum_{i=1}^N {h}_i s_i(t)
\; .
\end{equation} 
In the most realistic numerical experiments, the tapping is modelled
by a two-step dilation-relaxation process~\cite{mehta-barker} while
in more schematic models, the driving force is not explicitly
time-dependent~\cite{tetris0,alain,mauro,parking}. 
Two types of spatial dependence of the field will be considered below:
constant, ${h}_i = {h}$, for all sites $i$, and random 
$\overline{{h}_i {h}_j} = h^2 \delta_{ij}$.
There are however no physical differences between the two situations, since
a spatially constant field is as decorrelated from the ground states
of the Hamiltonian $H_J[{\bf s}]$ as a random one.

\subsection{Dynamical equations}
 
The dynamics of the spherical version of the model 
is better analyzed in terms of the autocorrelation function
$C(t,t') \equiv \sum_i \overline{\langle s_i(t)s_i(t') \rangle} /N$ 
and the linear response function
$R(t,t') \equiv 
\sum_i \overline{\langle s_i(t) \eta_i(t')}
 \rangle /(2TN)$ since, in 
the thermodynamic limit $N \rightarrow \infty$, $C$ and $R$ verify closed 
Schwinger-Dyson equations which read, for $t>t'$,~\cite{Cuku}
\begin{equation}
\begin{aligned} 
\frac{\partial C(t,t') }{ \partial t} = &
\, 
 -z(t) C(t,t') +
\frac{p (p-1)}{2}\int_0^t d t''  C^{p-2}(t,t'') R(t,t'') C(t'',t') \\
&
+ \frac{p}{2} 
\int_0^{t'} d t''  C^{p-1}(t,t'')  R(t',t'') 
+
h^2 \cos(\omega t) \int_0^{t'} d t'' \cos(\omega t'') R(t',t'') 
\\
\frac{\partial R(t,t')  }{ \partial t} = &
\,  -z(t)  R(t,t') +  \frac{p (p-1)}{2}
\int_{t'}^t d t''  C^{p-2}(t,t'') R(t,t'') R(t'',t') \;,
\\
z(t) = & \; 
T +  \frac{p^2}{2}
\int_0^t d t''   C^{p-1}(t,t'') R(t,t'') +
h^2 \cos ( \omega t) \int_0^{t} d t''  \cos (\omega t'') R(t,t'')
\; .
\label{eqdynp=3}
\end{aligned}
\end{equation}
These integro-differential equations are complemented by the 
equal-times conditions
$C(t,t)=1$, $R(t^+,t)=1$, the symmetry of the correlation,
$C(t,t')=C(t',t)$, and causality, $R(t',t)=0$.
In deriving these equations, a random initial condition at time $t=0$ has
been used, which can be interpreted as an equilibrium 
configuration at infinite temperature. 
An infinitely fast quench towards the final temperature $T$ is performed at 
$t=0$ and the evolution continues at subsequent times in isothermal 
conditions. 
The energy density $e(t) \equiv N^{-1} \langle 
\overline{H_J[\boldsymbol{s}]} \rangle$ is related to the constraint $z(t)$
through
\begin{equation}
e(t)=\frac{1}{p} \; [T-z(t)] \;.
\end{equation}
In Eqs.~(\ref{eqdynp=3}), the oscillatory field 
has been chosen to be constant in space.

In the $p=2$ case, the dynamics simplifies considerably since it 
can be solved directly from the Langevin
equation (\ref{langevin}). Indeed, this set of $N$ differential  
equations is diagonalized by using the  
basis of eigenvectors of the symmetric 
matrix $J_{i j}$. Denoting ${\boldsymbol \mu}$ the eigenvector
associated to the eigenvalue $\mu$, and $s_\mu \equiv {\boldsymbol
\mu} \cdot {\boldsymbol s}$  and 
${h}_\mu \equiv {\boldsymbol
\mu} \cdot{\boldsymbol {h}}$ the projections of the spin and field 
onto the eigenvectors, one obtains
\begin{equation}
\frac{\partial s_\mu (t)}{\partial t} = (\mu-z(t))s_\mu(t) + {h}_\mu 
\cos(\omega t)+\eta_\mu(t) \; .
\label{langevin2}
\end{equation}
Here we have considered, for convenience, a spatially uncorrelated 
random field.
The autocorrelation and response functions become
\begin{equation}
\begin{aligned}
C(t,t')= &\int_{-2}^2 d \mu \, \rho(\mu) \, \langle s_\mu(t) s_\mu(t') 
\rangle
\; , \\ 
R(t,t') =& \; \frac{1}{2T}
\int_{-2}^2 d \mu \, \rho(\mu) \, \langle s_\mu(t) \eta_\mu(t') \rangle
\; ,
\label{eqdynp=2}
\end{aligned}
\end{equation}
where $\rho(\mu) \equiv \sqrt{4-\mu^2}/2\pi$ for $\mu\in[-2,2]$, and zero 
otherwise, is the density of eigenvalues
of the random matrix $J_{i j}$.
The spherical condition reads $C(t,t)=1$ and after projection,
the amplitude of the field ${h}_\mu$ is random with zero mean and variance
$\overline{{h}_\mu {h}_{\mu'}} = {h}^2 \delta_{\mu\mu'}$. 

\subsection{Fluctuation-dissipation relation and effective 
temperatures}

We shall be interested in the fluctuation-dissipation
theorem ({\sc fdt}) and its possible modifications.
For driven systems, we do not expect this relation to be satisfied.
In equilibrium, {\it any} 
correlation function $C(t,t')=\langle O(t) O(t')\rangle$ 
[we assume, without lose of generality, that $\langle O(t) \rangle =0$],
and its associated linear response 
$R(t,t')=\delta\langle O(t)\rangle/\delta f(t')|_{f=0}$,
where the perturbation $f$ 
modifies the Hamiltonian  of the system according to 
$H \to H-f O$, satisfy the {\sc fdt}
\begin{equation}
R(t,t') = \frac{1}{T} \frac{\partial C(t,t')}{\partial t'} 
\;\;\;\;\;\;\;\;\;\;\;\;
t\geq t'
\;.
\label{FDT}
\end{equation}
For non-equilibrium systems, a possible extension is~\cite{Cuku,Cukupe} 
\begin{equation}
R(t,t') = \frac{1}{T_{\sc{eff}}(t,t')} \frac{\partial C(t,t')}{\partial t'}
\;\;\;\;\;\;\;\;\;\;\;\;
t\geq t'
\; .
\label{XFDT}
\end{equation}
Naively, this equation is simply a definition of the two-time function 
$T_{\sc eff}(t,t')$. 
This extension becomes non-trivial when one realizes
that, in the long waiting-time limit of many solvable models, 
this function only depends
on times via the correlation function itself,
\begin{equation}
T_{\sc eff}(t,t') = {\cal T}_{\sc eff}[C(t,t')]
\; .
\label{eq:Teff}
\end{equation}
By extension, it has been proposed 
that this exact result for solvable, mean-field like models, will also
apply to more realistic models with, {\it e.g.},
 finite range interactions.
Graphically, a convenient way of checking this ansatz is
to represent the integrated response function
$\chi(t,t') \equiv \int_{t'}^{t} d t'' R(t,t'')$ as a function
of $C(t,t')$ at fixed $t'$ and 
parametrized by the time difference $t-t'$ \cite{Cuku2}.
Equation~(\ref{eq:Teff}) 
implies that a master curve (i.e. independent of $t'$)
$\chi(C)$ is asymptotically reached.

In the aging case, three families of models 
have been found~\cite{review_aging}: (1) glassy models,
for which the $\chi(C)$ curve is a broken straight line 
with a first piece, from $C=1$ to 
$C=q_{EA}\equiv \lim_{t\to\infty}\lim_{t'\to\infty} C(t,t')$, 
of slope $-1/T$ and 
a second piece, from $C=q_{EA}$ to $C=0$, of slope $-1/T_{\sc{eff}}$, 
$T_{\sc{eff}}$
being finite; (2) `domain growth models', for 
which the $\chi(C)$ curve is still a broken straight line 
with the second piece having $T_{\sc{eff}}=\infty$;
(3) spin-glass models, for which 
the $\chi(C)$ curve has a straight line piece from  $C=1$ to $C=q_{EA}$
and a curved piece from $C=q_{EA}$ to $C=0$. 

The modification of these
plots in a system driven by shear-like forces has been studied
at the mean-field level in Refs. \cite{Cukulepe,BBK}. 
Numerically, the same behavior has been found in a sheared
supercooled liquid~\cite{BB}.
Finally, Langer and Liu studied a sheared foam and 
studied the {\sc fdt} by comparing
the fluctuations of the stress on the boundary and the 
corresponding compression modulus~\cite{liu}. 

At the theoretical level, it was shown in Ref.~\cite{Cukupe}
that the factor $T_{\sc{eff}}$ is indeed a {\it bonafide} temperature:
it is commonly called `effective temperature'.
A very appealing connection of this factor with the ideas of 
Edwards~\cite{edwards}, in the context of granular matter, 
and Stillinger-Weber~\cite{Stwe} 
in the context of glass forming liquids are currently being 
explored~\cite{inherent,Baetal,nicodemi2}. 
We shall come back to this point below.

Finally we wish to recall a rigorous bound that controls the maximum 
deviation  from {\sc fdt} that can be observed in an out of equilibrium 
system with Langevin dynamics~\cite{Cudeku}. 
When a time-dependent force of period $\tau$ is applied on 
the system the bound takes the form
\begin{equation}
\int^{t_w+\tau}_{t_w} d s \; 
\left(\frac{\partial C(t,s)}{\partial s} - T R(t,s) 
\right)
\leq \sqrt{ \left( {\cal N} \int^{t_w+\tau}_{t_w} d s \; D^2(t,s) \right) 
\left( \frac{W}{N} \right) }
\end{equation}
where $W$ is the total work done by the external force ${\boldsymbol{h}}$
on the system, per period, 
$W\equiv \int^{t+\tau}_t d s \; \langle {\boldsymbol v}(s) \cdot 
{\boldsymbol{h}}(s) \rangle$,
${\cal N} $ is a system dependent numerical factor and $D(t,s)$ is a
two-time correlation that in most cases of interest 
is again bounded by a numerical constant. In particular for
the model we treat in this paper, both ${\cal N} $ and $D^2(t,s)$
equal one.
Hence, for this model we have
\begin{equation}
\int^{t_w+\tau}_{t_w} d s \; 
\left(\frac{\partial C(t,s)}{\partial s} - T R(t,s) \right)
\leq \sqrt{ \tau \left( \frac{W}{N} \right) }
\; .
\label{bound}
\end{equation}

\subsection{Zero-drive dynamics: Jammed states and Reynolds dilatancy}

In the absence of a driving force, the model has a dynamical transition
at a ($p$-dependent) critical temperature $T_c$. 
For instance,
$T_c=1$ for $p=2$ and $T_c \simeq 0.6123$ for $p=3$.
Above $T_c$, the equilibration time 
is finite, and the system reaches equilibrium.
In this case, both time-translation invariance ({\sc tti}) and 
{\sc fdt} are
satisfied and Eqs.~(\ref{eqdynp=3}) reduce to the mode-coupling
equation for the so-called $F_{p-1}$
model~\cite{review_aging,Kith,Gotze}. 
The relaxation time diverges at $T_c$, and below $T_c$, the system does 
not equilibrate with its environment.
A quench from the high temperature phase towards the low temperature phase
is followed by the aging dynamics described in
the introduction: the relaxation time increases with $t_w$, which means 
that {\sc tti} is lost.
The {\sc fdt} is modified in the way described in the previous section.
One-time quantities have a slow relaxation 
--- typically power laws and, for instance,  
the energy density converges to a value
$e(t\rightarrow \infty) = e_{th}(T)$ which is higher than the equilibrium
value $e_{eq}(T)$.

From the static point of view~\cite{Crso,muchstudied}, 
$T_c$ is the temperature below which the energy landscape 
becomes fractured into many metastable states.
`Many' means that their number ${\cal N}(T,e)$ is such that
the thermodynamic limit
\begin{equation}
\lim_{N \rightarrow \infty} \frac{\ln {\cal N}(T,e)}{N}
\label{eq:11}
\end{equation}
exists and is finite.
The so-called complexity (or `configurational entropy')
$\Sigma(T,e) \equiv \ln {\cal N}(T,e)$ is hence {\it extensive}
in a range $e \in [e_{eq}(T),e_{th}(T)]$.
The effective temperature $T_{\sc{eff}}$
defined in the previous Section from the {\sc fdt} violations
can be computed directly from the dynamical equations.
A remarkable result is that it is also given by~\cite{Monasson}
\begin{equation}
\frac{1}{T_{\sc{eff}}} = \frac{\partial \Sigma(T,e)}{\partial e}
{\Big \vert}_{e=e_{th}}
\label{Teff}
\end{equation}
at $T=0$. [At finite temperature the free energy density replaces the 
energy density
in Eqs.~(\ref{eq:11}) and (\ref{Teff}).]
This relation is very similar to the thermodynamic
definition of temperature and
to the definition of Edwards' compactivity in granular systems.
We shall come back to these similarities in Section~\ref{conclusion}.

The complexity is a useful quantity to understand, for example, the 
difference between the dynamics of the $p=2$ and $p\geq 3$ models. 
For $p = 2$, one has $e_{eq} = e_{th}$, and the complexity
is irrelevant, $\Sigma =0$. 
It follows that $T_{\sc{eff}}= \infty$.
This case falls in the `domain growth family' defined in the 
previous subsection. Instead, if $p\geq 3$, $e_{eq} < e_{th}$,
$\Sigma >0$ in the region $[e_{eq},e_{th}]$, and $T_{\sc eff}$ is finite.

If, instead of a quench from a high temperature at $t=0$,
the dynamics starts from one of the metastable states with
energy density $e_{eq}<e<e_{th}$,
the system may escape this state, but 
in a time scale which diverges with $N$.
Ergodicity is broken, and the dynamics is blocked.
We could call these metastable states the `jammed states' 
of the system.
(They are known as the {\sc tap} states in the spin glass 
literature~\cite{muchstudied}.)

When a driving force is applied to the system, the dynamics 
also depends on the initial state. 
The driven dynamics following a quench under shear-like forces has been
investigated in Refs.~\cite{Cukulepe,BBK}.
When starting from a low-lying state, and {\it for a small enough
driving force}, the system remains trapped (jammed).
Only by applying a large force will the system escape the deep state
and have a dynamics like the one following a quench~\cite{Cukulepe}.
Hence, the system becomes able to move only by raising its
energy density: this is, in the glassy context, the Reynolds dilatancy 
effect~\cite{reviews,jorge}.

The effect of these trapping states on the dynamics
following a quench are beyond the mean-field approximation,
Eqs.~(\ref{eqdynp=3}). 
We propose to investigate this interesting aspect
by simulating in Section~\ref{beyond} 
 the Ising version of the model keeping {\it finite}
the number $N$ of interacting spins, so that jammed states
will be dynamically accessible by the system.  

\section{Driven spherical Sherrington-Kirkpatrick model}
\label{drivenp=2}

In this Section we study analytically the driven dynamics of the $p=2$ case.

\subsection{Existence of a dynamic transition}

Let us first study the $p=2$ case at $T=0$.
The Langevin equation~(\ref{langevin2}) can be integrated out; this 
yields
\begin{equation}
s_\mu (t) = \frac{e^{\mu t}}{\Omega(t)} \left[ 1 + {h}_\mu
 \int_0^t d t' \cos(\omega t')
e^{- \mu t'} \Omega(t') \right],
\end{equation}
where the function 
\begin{equation}
\Omega(t) \equiv \exp \left[ \int_0^t d t' z(t') \right]
\end{equation}
has been defined.
The spherical constraint determines $\Omega(t)$ through the equation 
\begin{equation}
\label{omega}
\Omega(t)^2 = f(t) + h^2 \int_0^t d t' \int_0^t d t'' f \left( t-
\frac{t'+t''}{2} \right)
\Omega(t') \Omega(t'') \cos(\omega t') \cos (\omega t'')
\;,
\end{equation}
with $f(t)$ given by 
\begin{equation}
f(t) \equiv \int_{-2}^2 d\mu \, \rho(\mu) \, e^{2\mu t}
\; .
\end{equation}
The autocorrelation and response functions, together
with the energy density, read
\begin{equation}
\begin{aligned}
C(t,t_w) =& \;\frac{1}{\Omega(t) \Omega(t_w)} \left[
f \left( \frac{t+t_w}{2} \right) 
\right.
\\
&
+
\left.
 h^2 \int_0^t d t' \int_0^{t_w} d t'' \cos(\omega t') 
\cos(\omega t'') \, \Omega(t') \Omega(t'') f \left( \frac{t+t_w-t'-t''}{2}
 \right) \right] \; , \\
R(t,t_w) = & 
\;\frac{\Omega(t_w)}{\Omega(t)} f\left( \frac{t-t_w}{2} \right)
\; , \quad
e(t) =   \,
- \frac{z(t)}{2}=-\frac{1}{2} 
\frac{d \ln[\Omega(t)]}{d t}
\; .
\label{p=2dyn}
\end{aligned}
\end{equation}
Equation~(\ref{omega}) determines $\Omega(t)$ and, consequently,
all dynamic quantities.
Unfortunately, 
it cannot be completely solved analytically 
except in some asymptotic limits.

Two special cases have been studied previously, namely the zero field
behavior of the model, and the case of a dc field.
In the former, as discussed in Section~\ref{definition},
 there is no finite equilibration time below $T_c=1$ and the 
system ages forever~\cite{p=2,cude}. 
In the presence of a dc magnetic field of amplitude $h$ \cite{cude2}, 
Eq.~(\ref{omega}) yields $\Omega(t) \sim \exp( \lambda^0(h) t)$, where
$\lambda^0(h) \equiv (2+h^2)/ \sqrt{1+h^2}$. 
The main effect of the field is to introduce a new time scale 
${t_r}^0(h) \equiv [\lambda^0(h)-2]^{-1}$ in the problem.
For times $t_w \ll {t_r}^0(h)$ after the quench, aging is observed 
as in the zero-field case,
whereas at later times $t_w \gg {t_r}^0(h)$, the system has reached
its equilibrium state in a field: aging is interrupted.
The energy density $e(t)$ converges to 
$e_\infty (\omega=0,h)= -(2+h^2) / (2 \sqrt{1+h^2})$.  

\begin{figure}
\begin{center}
\psfig{file=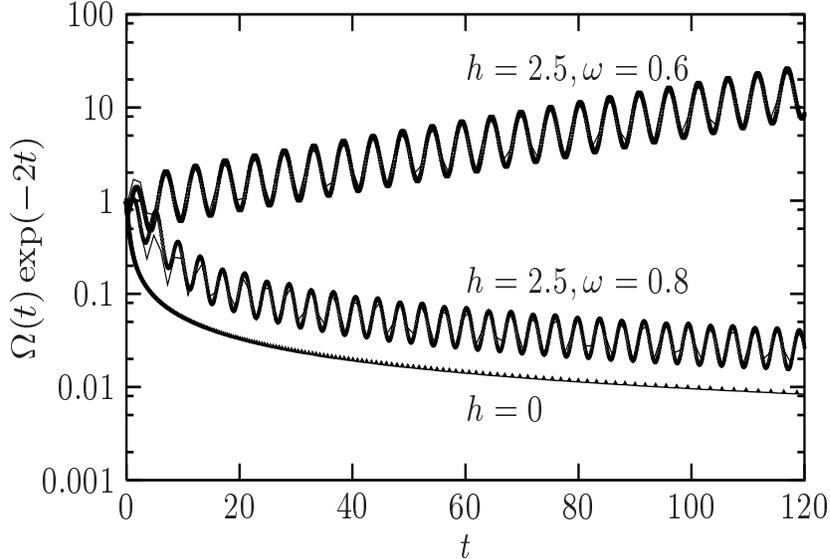,width=11cm,height=7.5cm}
\caption{The function $\Omega(t)\exp (-2t)$ in the three cases $h=0$ and
$h=2.5$ with $\omega=0.8$ (glassy phase)
 and 0.6 (liquid phase), from bottom to top. 
The points are the direct integration of 
Eq.~(\ref{omega}), while the full lines are fits
to Eqs.~(\ref{num}).
 Except at very short times, the approximate 
forms in Eqs.~(\ref{num}) yield the correct behavior.}
\label{Omega}
\end{center}
\end{figure}

At finite $\omega$, the numerical solution of Eq.~(\ref{omega}) displayed 
in Fig.~\ref{Omega} suggests
the following asymptotic behavior:
\begin{equation}
\begin{aligned}
h < h^\star (\omega) \quad\quad & \Omega(t) \sim  c \, 
\frac{e^{2t}}{ t^{3/4} } ( b \cos(2\omega t + \phi) + 1) \; , \\
h > h^\star (\omega) \quad\quad  & \Omega(t) \sim c' \, 
\exp \left[ \lambda(\omega,h)t
+ a\cos( 2 \omega t + \phi) \right] \; , 
\label{num}
\end{aligned}
\end{equation}
where $c$, $c'$, $a$, $b<1$ and $\phi$ are numerical constants, and with
$\lambda(\omega,h) > 2$.
This means that if the amplitude of the driving force is sufficiently small,
the ac field does not introduce a new time scale
and a full aging behavior is observed.
For stronger amplitudes of the field, a time scale defined by 
\begin{equation}
t_r(\omega,h) \equiv [\lambda(\omega,h)-2]^{-1}
\label{time-scale}
\end{equation}
 is generated and the
relaxation time becomes finite: aging is stopped
by the driving force.
There is hence a critical field $h^\star$ separating 
these two different regimes.
It is clear from Fig.~\ref{Omega} 
that the agreement between Eqs.~(\ref{num}) and 
the numerical solution is very good, after a short transient.

In the following we characterize more precisely 
these two phases as well as the
transition between them.
 
\subsection{The transition line in the plane $\boldsymbol{(\omega,h)}$}

We have seen that, at fixed pulsation $\omega$, there exists a 
well-defined transition line
where the relaxation time in an ac field diverges, allowing to
distinguish between a `glassy' and a `liquid' state.
We anticipate the discussion of the last section to emphasize 
that this transition is a non-equilibrium phase transition 
which is hence of a different nature that the transition
taking place at $T_c$ in the absence of the driving force.

The transition line $h^\star(\omega)$
may be understood and estimated from a simple physical argument.
At $\omega=0$, the  dc field introduces a finite relaxation time
${t_r}^0(h)$.
The most naive requirement  for the system to keep a finite relaxation time 
in an ac field of period $\tau$ is given by
${t_r}^0(h) \lesssim \tau$.
The transition line is then estimated by the relation
\begin{equation}
{t_r}^0(h^\star) \simeq \tau 
\;\;\;\;\;
\Leftrightarrow
\;\;\;\;\;
\omega \propto \left[\frac{2+{h^\star}^2}
{\sqrt{1+{h^\star}^2}} -2 \right].
\label{heuristic}
\end{equation} 

\begin{figure}
\begin{center}
\psfig{file=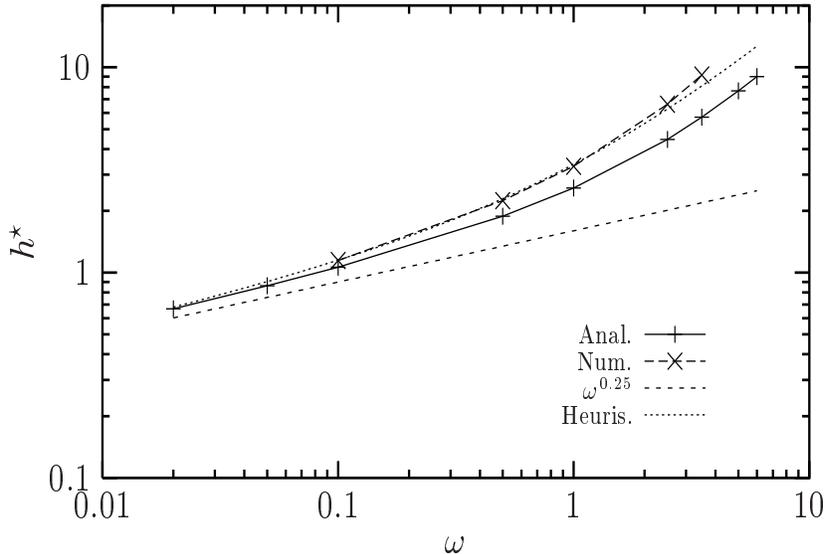,width=11cm,height=7.5cm}
\caption{The transition line $h^\star (\omega)$, estimated by analytical,
numerical and heuristic tools, together with the small
frequency behavior $h^\star \propto \omega^{0.25}$.}
\label{transition}
\end{center}
\end{figure}

A more refined computation can also be performed. 
Using the fact that,
when $h>h^\star$, 
$\Omega(t)$ is given asymptotically by Eq.~(\ref{num}) with 
$\lambda>2$ allows us to 
neglect the term $f(t)$ in the right hand side of Eq.~(\ref{omega}); 
thus
\begin{equation}
\Omega(t)^2 \simeq h^2 \int_{-2}^{2} d \mu \rho (\mu) e^{2\mu t}
\left[ \int_0^t d t' \cos (\omega t') \Omega (t') e^{-\mu t'}
\right]^2 .
\end{equation}
Inserting the form $\Omega(t) = \exp(\lambda t +B(t))$, and formally
integrating by parts yields
\begin{equation}
e^{B(t)}= h^2 \int_{-2}^{2} d \mu \rho (\mu) \left[
\Re \left( e^{i \omega t} \sum_{k=0}^{\infty}  \frac{(-1)^k}  
{(\lambda(\omega,h) - \mu +i \omega)^{k+1}}\cdot 
\frac{d^ke^{B(t)}}{d t^k} 
\right) \right]^2.
\end{equation}
Since the function $B(t)$ is periodic with angular velocity $2\omega$, 
this corresponds to a development in powers of $\omega$. In the small
frequency limit, the equation can be closed by keeping only the leading terms
of the development, and this yields a relation between 
$(\lambda,\omega,h)$:
\begin{equation}
\frac{\pi}{h^2} = \int_{-1}^1 dx \frac{\sqrt{1-x^2}}
{(\lambda-2x)^2+\omega^2} \;, 
\label{lambda}
\end{equation}
from which we estimate the transition line where the time scale 
$t_r(\omega,h)$ given in Eq.(\ref{time-scale}) diverges.
This is
equivalent to the condition $\lambda(\omega,h^\star)=2$. 
It is easily shown that in the small frequency limit, the scaling
${t_r}^0 (h^\star) \simeq \tau$ is recovered, in accordance with
our heuristic argument, Eq.~(\ref{heuristic}).
In Fig.~\ref{transition}, the analytic estimate for the 
transition line $h^\star (\omega)$ is compared with
its direct numerical estimation, and with 
our heuristic argument. 
Also plotted is the small frequency behavior
of the critical field, $h^\star \propto \omega^{0.25}$.

Moreover, Equation~(\ref{lambda}) implies 
\begin{equation}
\lim_{\omega \rightarrow 0} \lambda(\omega,h) = 
\lambda^0 \left( \frac{h}{\sqrt{2}} \right),
\end{equation}
which means that, in the limit $\omega \rightarrow 0$, the field 
acts as a constant field with an amplitude given by its root mean square
 value, which is physically reasonable.
Importantly enough, this means that the effect of a dc field with 
$\omega$ strictly zero and the limit of an ac field with vanishing frequency 
are different. In the case $p=2$ this feature has no effect in the 
value of the transition field $h^\star$ since it vanishes in both cases.
However, when $p\geq 3$ we shall find a non trivial consequence 
of this result (see Section~\ref{drivenp=3}).

\subsection
{\bf Below the transition ${\boldsymbol{h < h^\star(\omega)}}$: 
aging in an ac field.}

Below the transition, Eq.~(\ref{num}) shows
that the energy density slowly converges towards
its asymptotic value $e_\infty (\omega,h<h^\star) = -1$ 
as a power law $e(t) - e_\infty \sim t^{-1}$
with a superimposed oscillation at a frequency $\omega/\pi$.
This asymptotic value is independent of $h$. 
A slow (here a power law) convergence of one time quantities
is typical of aging systems~\cite{review_aging}.
The angular velocity  $2\omega$ is expected since Eq.~(\ref{langevin2}) 
remains unchanged
by the transformation $t\rightarrow t+ \pi/\omega$; 
$s_\mu \rightarrow - s_\mu$; $z \rightarrow z$.
The constraint $z(t)$, and the energy density, are
then $\pi/\omega$-periodic.  

\begin{figure}
\begin{center}
\psfig{file=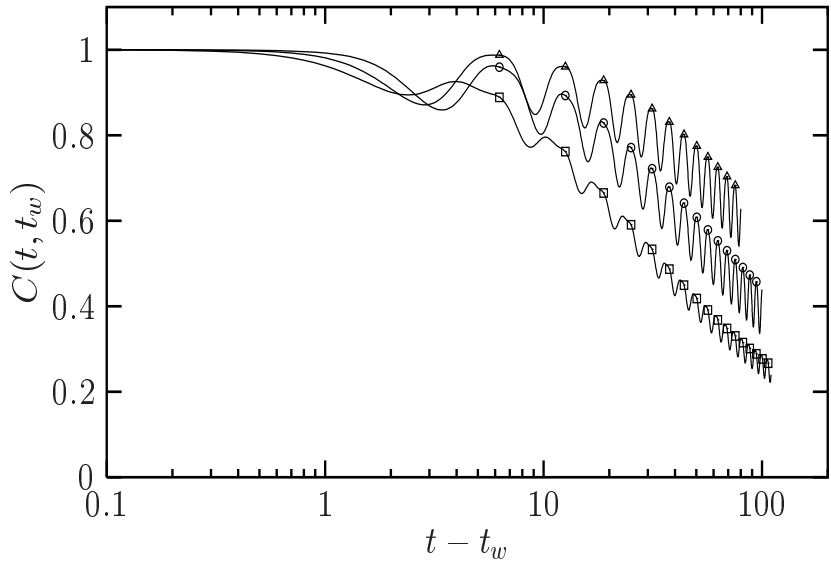,width=11cm,height=7.5cm}
\caption{Correlation function of the $p=2$ model at $T=0$ 
for $h=0.5$, $\omega=1$  and waiting times
$t_w=5$, $10$, $20$, from bottom to top.
The points fall on the discrete times 
$t=n\tau$, with $n$ an integer.}
\label{agingp=2}
\end{center}
\end{figure}

That {\sc tti} is also lost is demonstrated by looking at 
two-time quantities, typically correlation functions.
The behavior of $C(t,t_w)$ is represented in Fig.~\ref{agingp=2}.
This figure also  shows  the interesting feature that the 
dynamics can be decomposed into two well-separated time scales.
At short time separation $t-t_w$, the time-scale for the 
approach to  the plateau at $C=q_{\sc ea}$ (note that
$q_{\sc ea}=1$ at $T=0$)
does not depend on $t_w$, whereas the decay
from the plateau towards zero arises in a second time scale 
that clearly depends on $t_w$. During the waiting-time dependent decay the 
curves have oscillations. 
Quantitatively, the behavior of the correlation function 
is entirely dominated at long times by the term
$C(t,t_w) \sim f(\frac{t+t_w}{2})/ (\Omega(t)\Omega(t_w))$.
This implies that, apart from the oscillation, 
the correlation scales as $t/t_w$ for $t \gg t_w$; we have numerically
checked this point.

In tapping experiments, data are 
obtained for times of the form $t= n \tau$, where $n$ is an integer.
We have represented these data by points
in Fig.~\ref{agingp=2}, and the experimental measurements
would be very similar to the usual aging case.

\begin{figure}
\begin{center}
\psfig{file=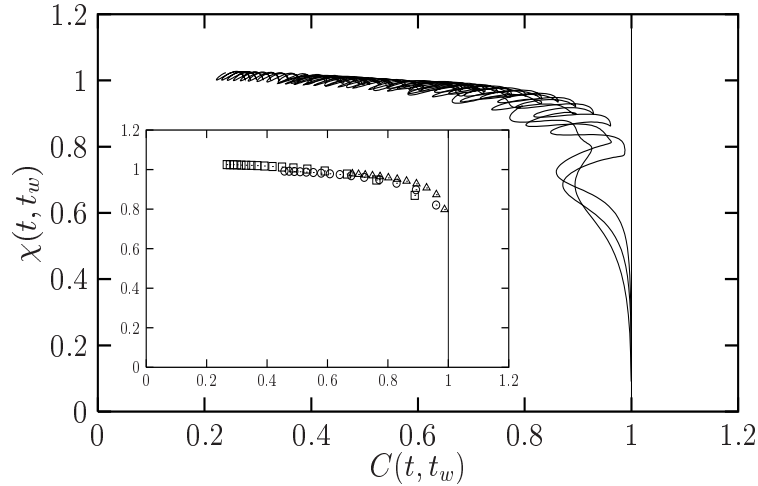,width=10cm,height=6.5cm}
\caption{Parametric plot of the integrated response-correlation 
of the $p=2$ model at $T=0$
for $h=0.5$, $\omega=1$  and waiting times
$t_w=5$, 10 and 20. 
For the {\sc fdt} to be satisfied, the points have to lie
on the vertical line.
The inset represents the same data, but for times $t=n\tau$; the symbols are
the same as in Fig.~\ref{agingp=2}.}
\label{respp=2}
\end{center}
\end{figure}

The  response function is analyzed in Fig.~\ref{respp=2},
where the parametric plot of the integrated response against the 
correlation is built.
As expected, there are strong violations of the {\sc fdt}.
These violations are however similar to those encountered
in the relaxational case, in the sense that there are 
two distinct behaviors, depending on the time scale considered.
For a time separation  $t-t_w \ll t_w$, 
{\sc fdt} is not satisfied, and no
effective temperature can be defined. 
For the second time scale $t \gtrsim t_w$ and longer, on the contrary, all 
points fall nicely on a straight line, with a superimposed oscillation.
The bound (\ref{bound}) shows however that
the deviation from {\sc FDT} can only be significant at large times,
when both $\partial C(t,t_w)/\partial t_w$ and $R(t,t_w)$, 
integrated over a period,
are small. This is confirmed by Fig.~\ref{respp=2}.

The inset of Fig.~\ref{respp=2} shows $\chi(C)$ for the discrete 
times $t=n\tau$. If one uses this stroboscopic measurement, it remains 
clear that {\sc FDT} modifications
are well accounted for by the ansatz (\ref{eq:Teff}).
Analytically, 
neglecting as above the second term in the correlation function, 
the effective temperature defined through Eq.~(\ref{XFDT}) scales 
as $T_{\sc{eff}} \sim t_w^{1/2} \rightarrow \infty$.
An infinite effective temperature 
is also  present in the relaxational dynamics of this model,
and is typical of  domain growth models.

With these simple considerations we have argued that the
effect of a small ac field does not change the aging behavior
of this model at $T=0$. This is to be confronted to the 
effect of non-Hamiltonian perturbations that 
change the aging scaling, as seen for instance in the 
correlation function that becomes 
$C(t,t_w) \sim f(\exp(\sqrt{t_w}-\sqrt{t}))$~\cite{Cukulepe}.
At non-zero temperature, shear-like forces  
introduce a finite relaxation time
after which the dynamics is stationary. 
We have not solved analytically the $p=2$ model at finite temperature 
in an ac field, 
but the  numerical solution 
of the full equations suggests that there is a finite region
of the phase diagram $(T,h,\omega)$ in which aging effects 
survive.
This explicitly shows that 
the effect of shear and tapping are rather
different.

\subsection{\bf Above the transition: driven steady state.}

Above the transition, the situation is simpler. 
After a time scale $t_r(\omega,h) = [\lambda(\omega,h)-2]^{-1}$ 
the slow dynamics is lost.
The relation (\ref{lambda}) implicitly determines $\lambda$, which
is an increasing function of the amplitude $h$
of the driving force.
In the stationary state, 
the energy density oscillates with an angular velocity $2 \omega$
around its asymptotic value $e_\infty(\omega,h>h^\star)=-
\lambda(\omega,h)/2$.  
The behavior of the correlation is dominated by the second term in Eq.
(\ref{p=2dyn}). This shows that it decays towards zero in 
a time scale of order $t_r(\omega,h)$: there is no more aging.

\section{Behavior for $\boldsymbol{p \ge 3}$}
\label{drivenp=3}

Going beyond the solution of the unperturbed case \cite{Cuku}
to solve the set of coupled integro-differential equations~(\ref{eqdynp=3}) 
analytically is a very hard task. In order to illustrate the main
 properties of
the solution, we solved Eqs.~(\ref{eqdynp=3}) numerically, by constructing the
two-time solution step by step 
in time.
We first demonstrate that also in the case $p \ge 3$, there
exists a transition line below which glassy
properties of the undriven model persist, and study 
then the {\sc fdt}-violations in both phases.

\subsection{Evidences for a dynamical transition}

Let us recall first the effect of a dc magnetic field of amplitude
$h$ on the 
system~\cite{Cavagna}. 
In contrast to
the spherical Sherrington-Kirkpatrick, the spin glass phase
may exist in a dc magnetic field. There are both a dynamic and a static 
transition between the spin glass and the paramagnetic phases.
At $T=0.2$, the dynamic transition takes place at 
${h}^\star(\omega=0) \simeq 1.1$~\cite{Cavagna}. 

\begin{figure}
\begin{center}
\psfig{file=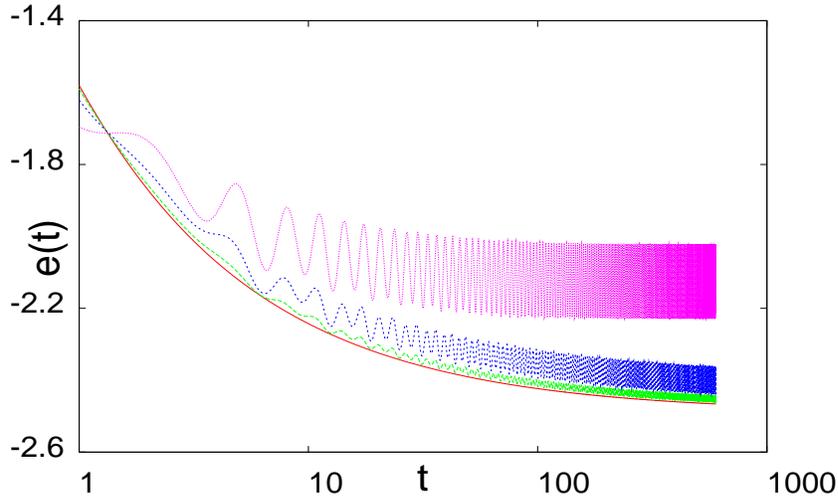,width=6.5cm,height=10cm,angle=-90}
\caption{Time dependence of the energy density under an ac field with 
amplitude $h=0,0.3,0.6,1$ (from bottom to top) 
and angular velocity $\omega=1$. The curve for 
$h=1$ approaches rapidly an asymptote, whereas $e(t)$ 
still evolves at large times for the other cases.}
\label{energy_T06}
\end{center}
\end{figure}

As in the case $p=2$, we begin our study by focusing on the
behavior of the energy density $e(t)$.
It is displayed in Fig.~\ref{energy_T06}
at temperature $T=0.6$ and angular velocity $\omega=1$, for different
field amplitudes. 
The long time behavior of $e(t)$ indicates that
the transition 
occurs for a field $0.6 < h^\star <1.0$ at this frequency. 
The asymptotic value of the energy density slightly increases below 
the transition, and reaches a much larger value, when $h > h^\star$.

\begin{figure}[t]
\centerline{
\hbox{\epsfig{figure=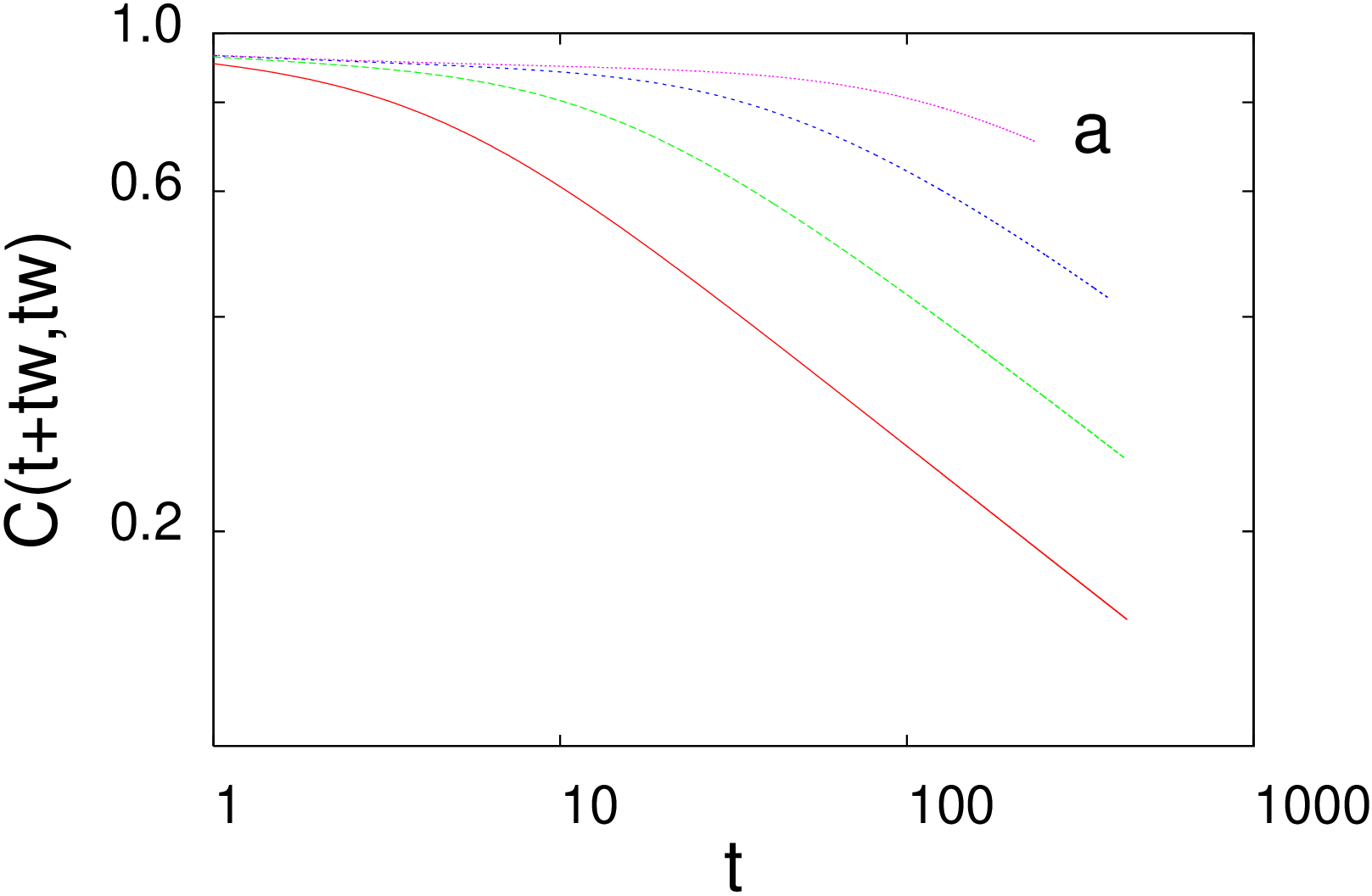,width=5cm,angle=-90}}
\hspace{.25cm}
\hbox{\epsfig{figure=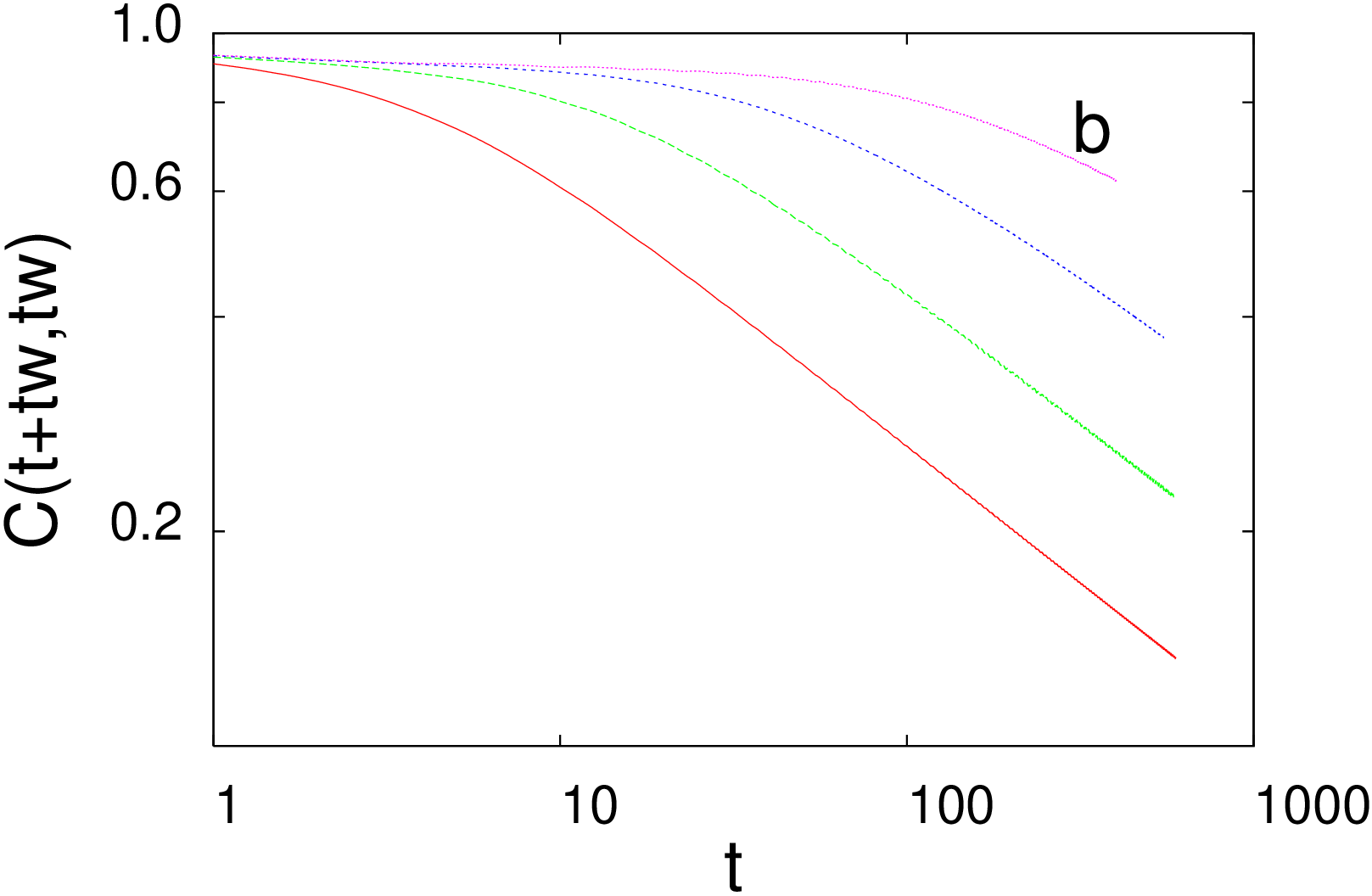,width=5cm,angle=-90}}
 }
\vspace{.25cm}
\centerline{
\hbox{\epsfig{figure=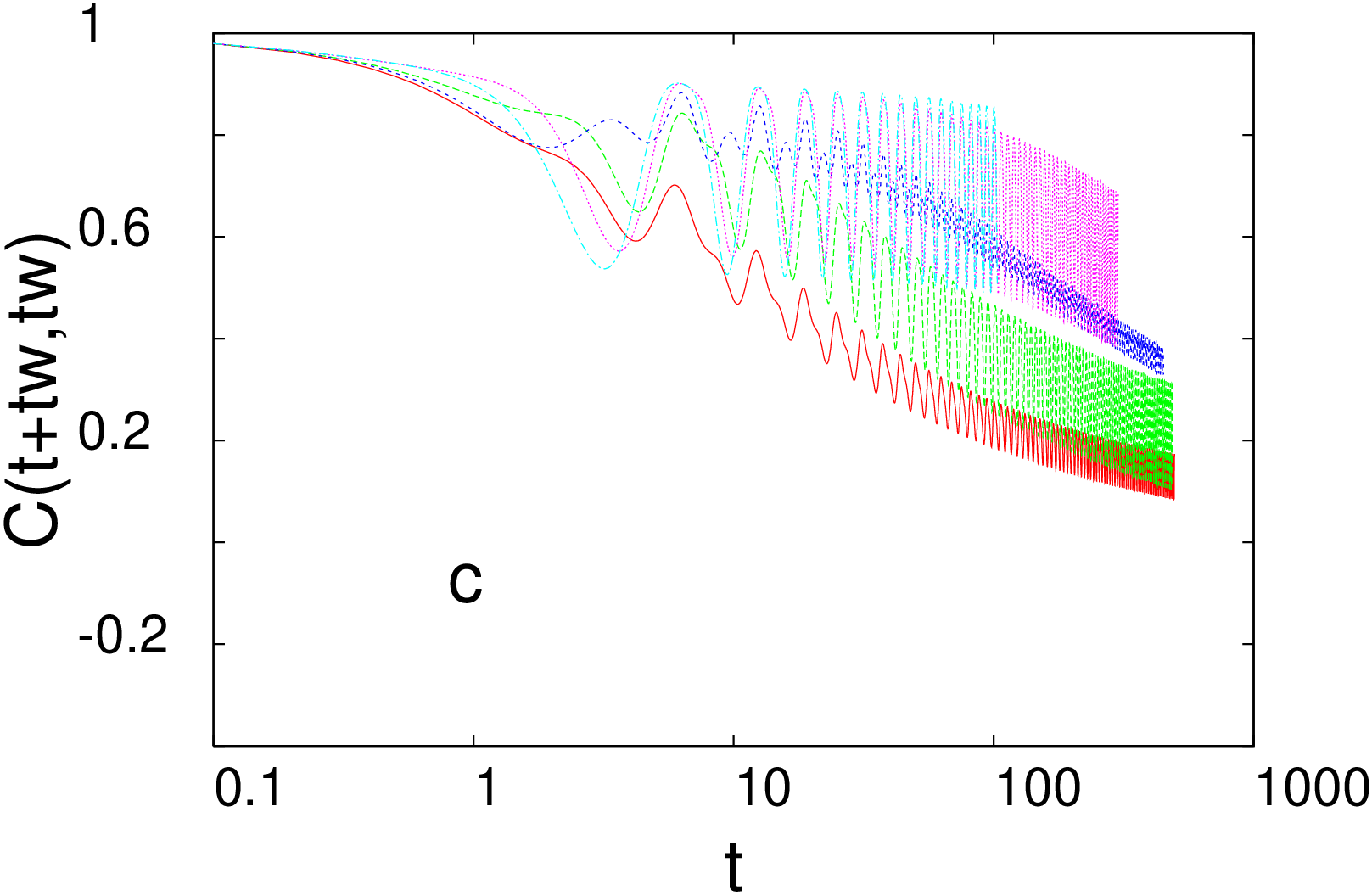,width=5cm,angle=-90}}
\hspace{.25cm}
\hbox{\epsfig{figure=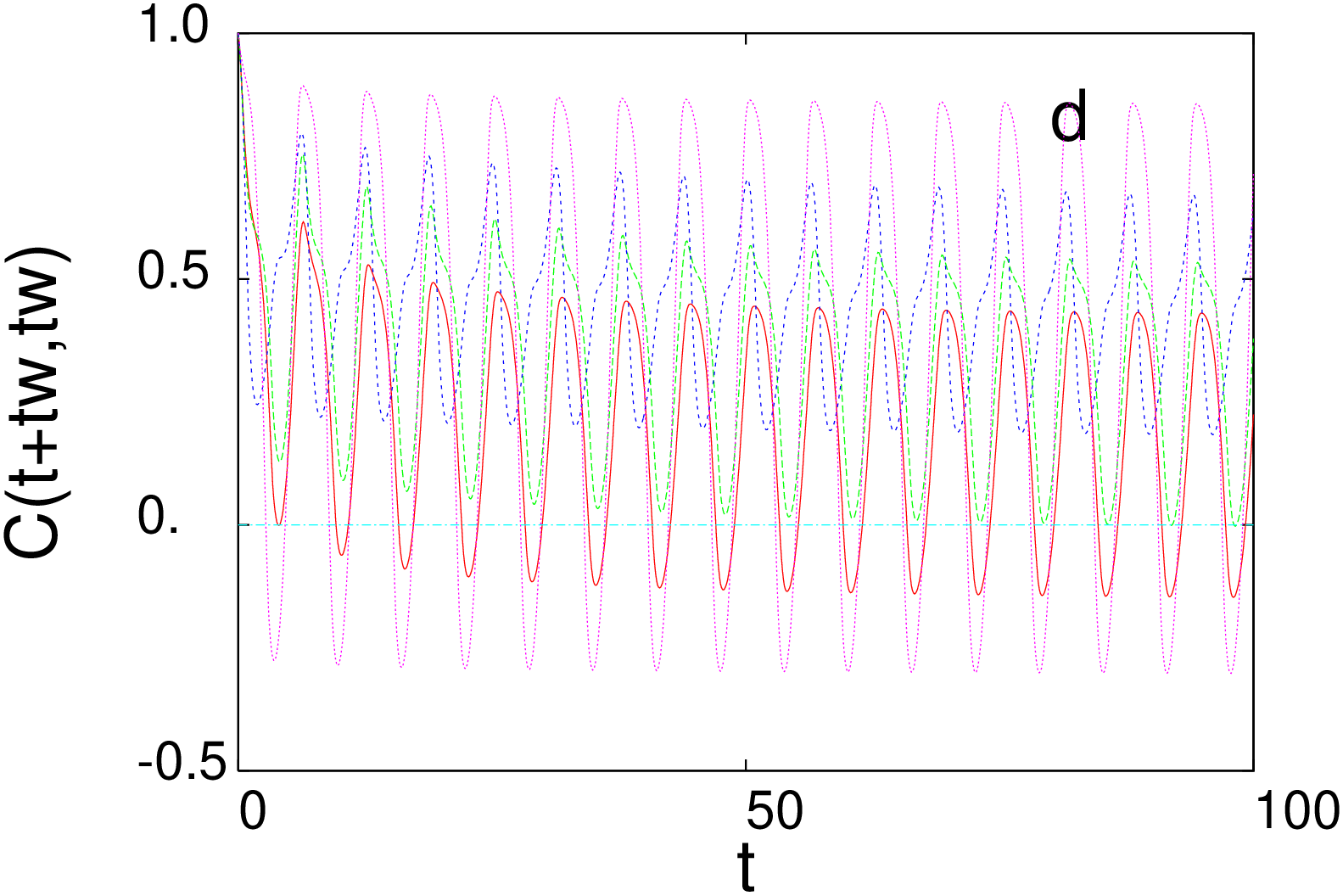,width=5cm,angle=-90}}
\vspace{.5cm} 
}
\caption{
Each panel shows the auto-correlation $C(t+t_w,t_w)$ as a 
function of $t$, for four 
values of the waiting-time $t_w=3,12,49,198$ at $T=0.2$. 
The angular velocity  of the 
applied field is $\omega= 1$. In panels a to c the 
scale is logarithmic, the amplitudes of the 
applied fields are $h=0,0.1,1$ and the system is in 
its glassy phase. In panel d the scale 
is linear,  $h=2$ and the slow dynamics is suppressed.} 
\label{T02-corr-hfdep}
\end{figure}

In the following we concentrate on the temperature 
$T=0.2 \simeq 0.32 \, T_c$, {\it i.e.}
well below the zero-field transition.
We focus first on the dependence of the two-time autocorrelation functions
on the amplitude of the applied field.
In Fig~\ref{T02-corr-hfdep}-a, the field is $h=0$, and we
observe the usual aging~\cite{Cuku}.
It is then clear
that aging is still present if the field is not too strong, $h=0.1$ and 
$h= 1$ in Figs.~\ref{T02-corr-hfdep}-b and c,
whereas at a stronger  field, $h=2$ in Fig.~\ref{T02-corr-hfdep}-d, 
the correlation very rapidly 
tends to zero, with a superimposed oscillation.
These observations reinforce the evidence in favour of
 a dynamic transition which, at $T=0.2$,
occurs at a field strength $1< h^\star < 2$, for 
$\omega =1$.

\begin{figure}
\centerline{
\hbox{\epsfig{figure=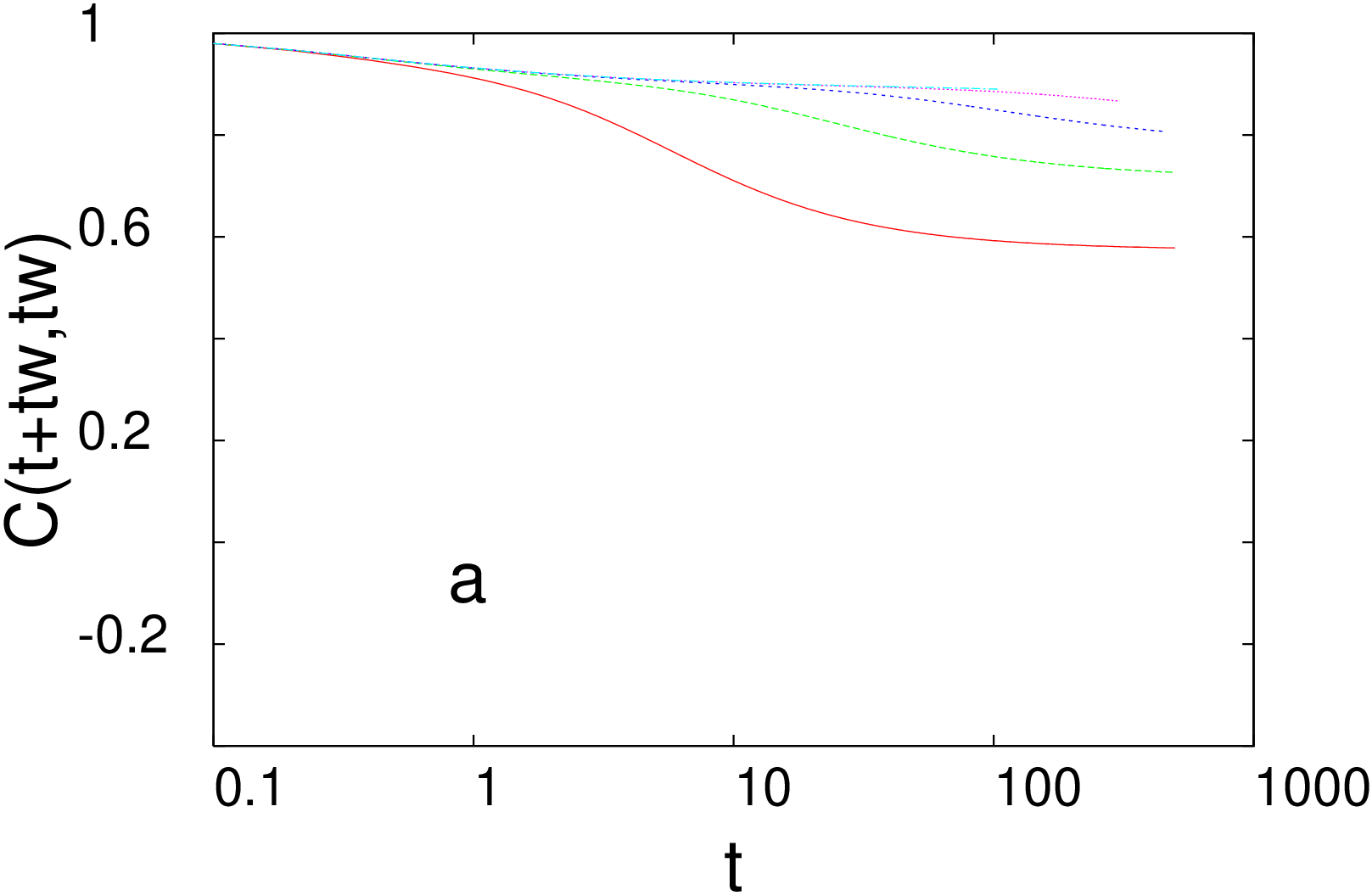,width=5cm,angle=-90}}
\hspace{.25cm}
\hbox{\epsfig{figure=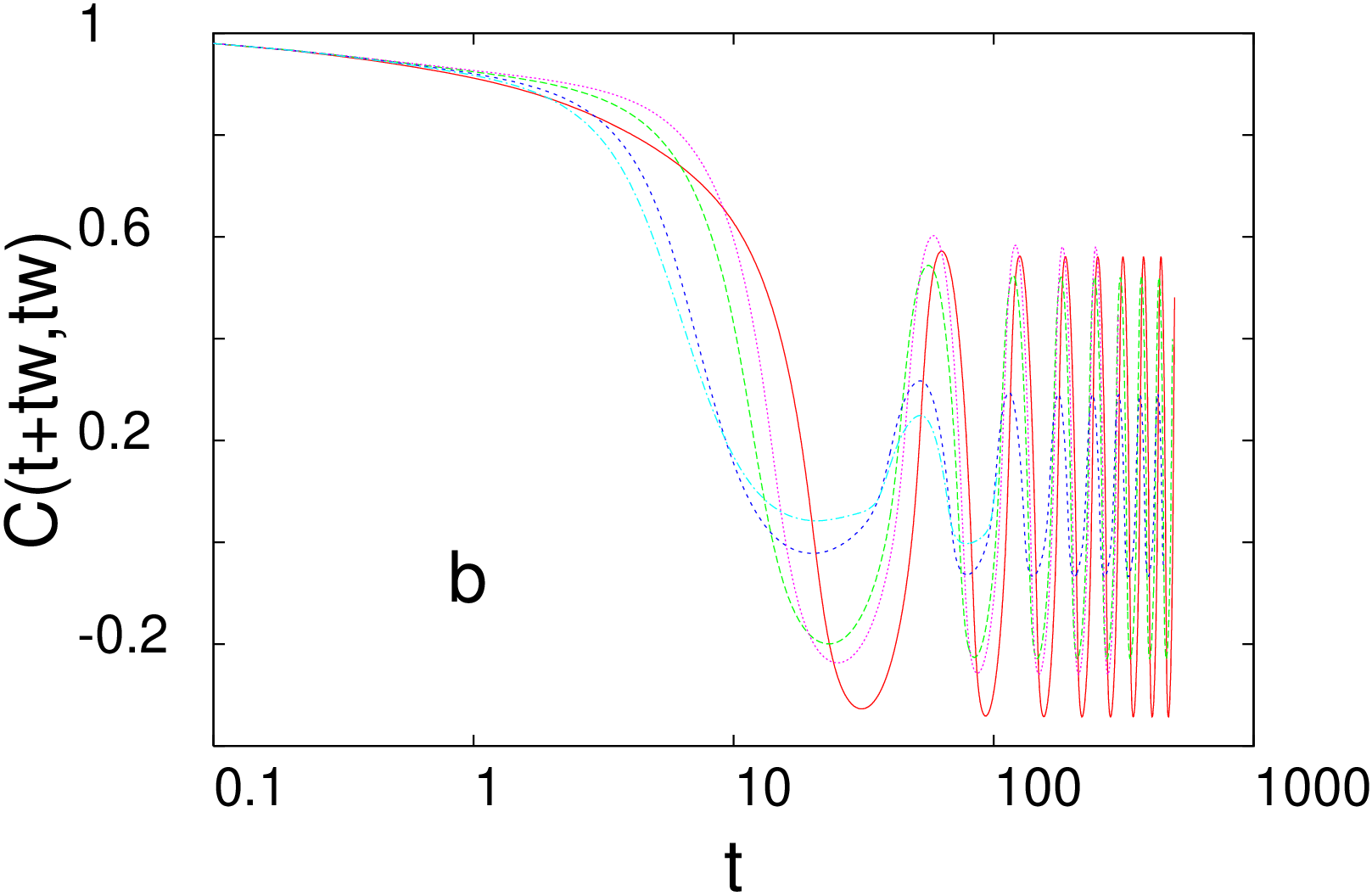,width=5cm,angle=-90}}
 }
\vspace{.25cm}
\centerline{
\hbox{\epsfig{figure=corr-w1-hf1-T02.ps,width=5cm,angle=-90}}
\hspace{.25cm}
\hbox{\epsfig{figure=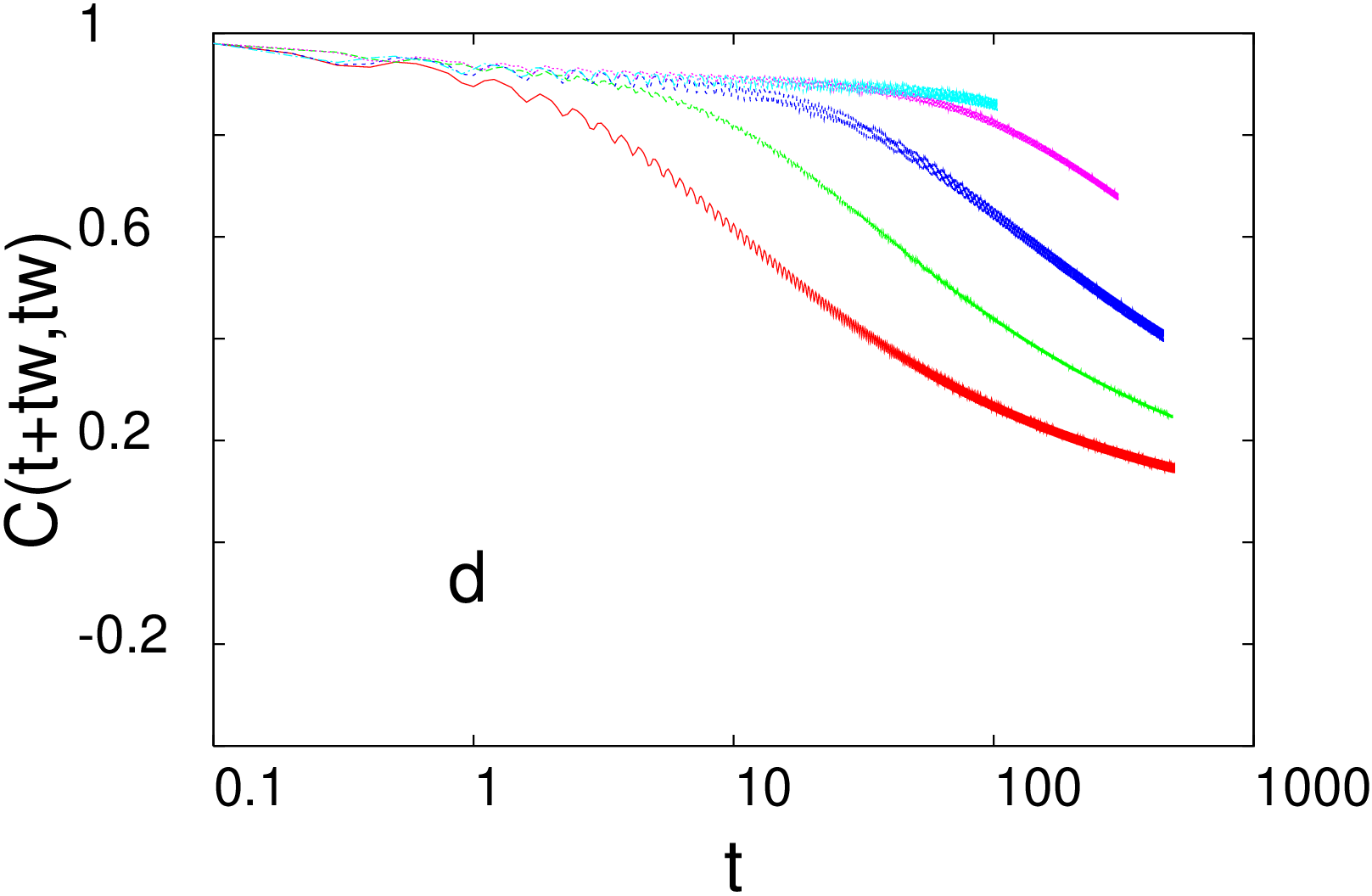,width=5cm,angle=-90}}
\vspace{.5cm} 
}
\caption{
Each panel shows the auto-correlation $C(t+t_w,t_w)$ as a 
function of $t$, for four 
values of the waiting-time $t_w=3,12,49,198$ at $T=0.2$. The strength of the 
applied field is $h=1$ and the frequencies are 
$\omega=0, 0.1, 1, 10$ from panel a to panel d.  
The transition in an ac-field of amplitude $h=1$, 
from the liquid to the glassy phase,   
occurs at a frequency $0.1< \omega_c < 1$. 
} 
\label{T02-corr}
\end{figure}

We turn now to the question of identifying the 
critical line in the $\omega$ direction.
Fig.~\ref{T02-corr} shows the auto correlation function at fixed
amplitude of the applied field and several values of the 
angular velocity.
In Fig.~\ref{T02-corr}-a, the frequency is zero,
and aging is observed with a two-step decay
of the correlation.
Note that the second decay is towards a value
$q_0 > 0$, as opposed to the zero-field case~\cite{Cavagna}.
The frequency in increased in Figs.~\ref{T02-corr}-b,c,d.
At intermediate frequency, Fig.~\ref{T02-corr}-b, 
aging is suppressed.
When $\omega$ is further increased, Figs.~\ref{T02-corr}-c,d, it is 
clear that the glassy phase  is  reentrant.

\begin{figure}
\begin{center}
\psfig{figure=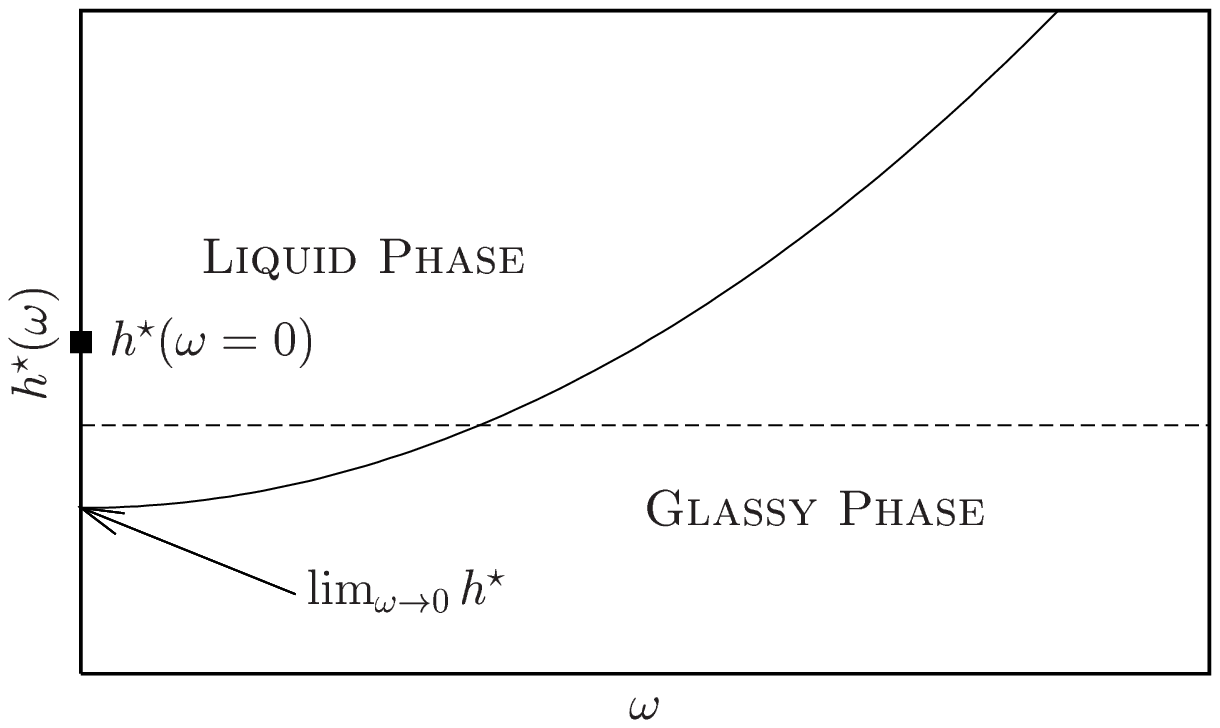,width=9cm}
\caption{The full line is a schematic
representation of the transition line $h^\star(\omega)$ for $p \ge 3$.
The angular velocity in the four panels in Fig.~\ref{T02-corr}
move from left to right along the horizontal dashed line 
(see the text for more details).}
\label{sketch}
\end{center}
\end{figure}

It is of course very difficult to determine 
the transition line  $h^\star(\omega)$ with accuracy, 
by the numerical solution of the dynamical equations. We are hence 
not able to draw a figure similar to Fig.~\ref{transition} for $p \ge 3$.
We show in Fig.~\ref{sketch} a schematic representation of this
critical line with the feature, already encountered for $p=2$, that
the limit $\omega \rightarrow 0$ is peculiar.
From the numerical solution at $\omega>0$ it seems that
\begin{equation}
\lim_{\omega \rightarrow 0} h^\star (T,\omega) < {h}^\star(T,\omega=0)
\; .
\end{equation}
This is represented in Fig.~\ref{sketch} and
explains the reentrance of the glassy phase
described in Fig.~\ref{T02-corr}, when the frequency increases.
Note however that we cannot numerically discard the possibility 
of a non-monotonous, but smooth, 
behavior of the critical line at $\omega<0.1$.

\subsection{Fluctuation-dissipation theorem}
 
\begin{figure}[h]
\centerline{
\hbox{\epsfig{figure=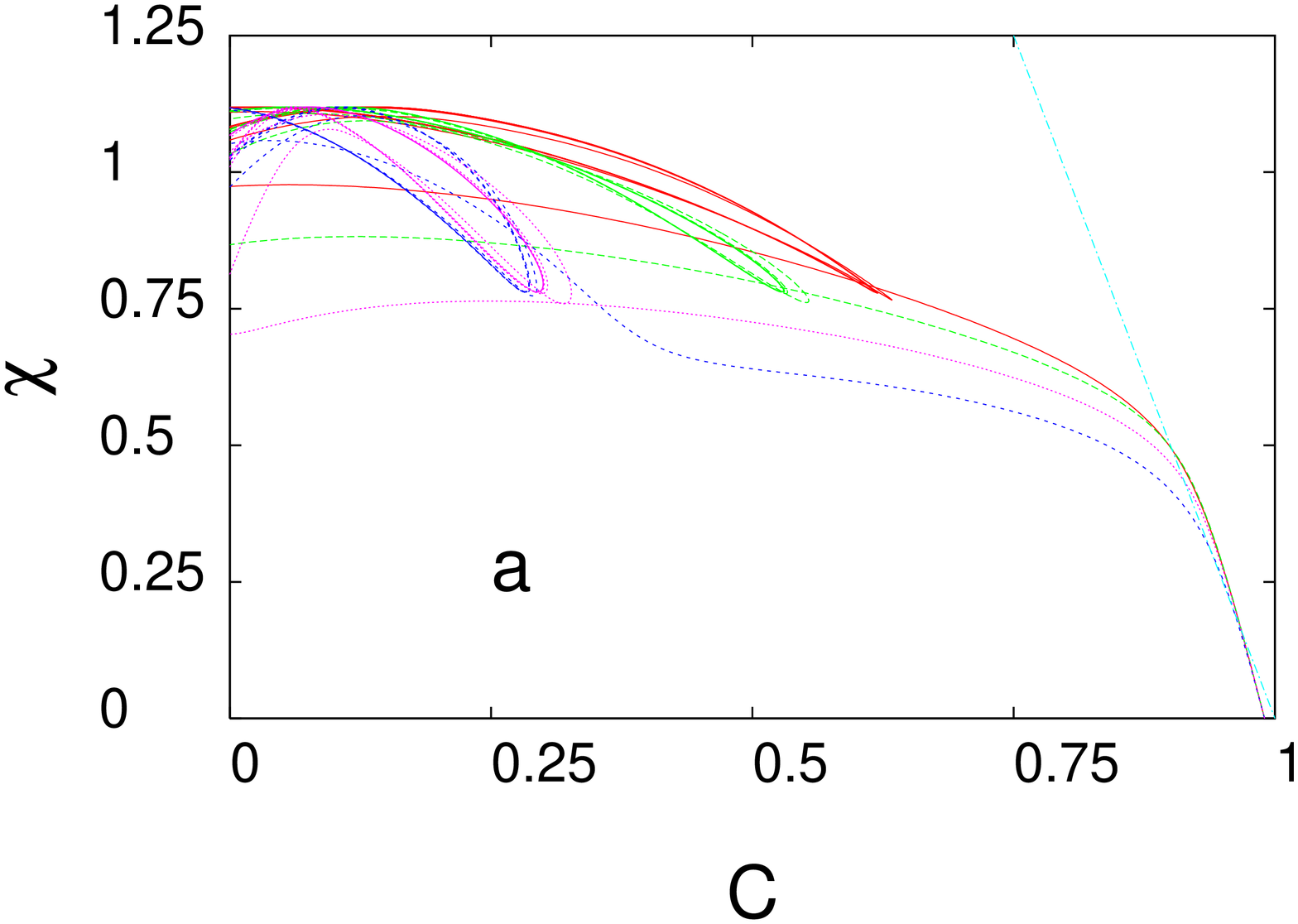,width=5cm,angle=-90}}
\hspace{.25cm}
\hbox{\epsfig{figure=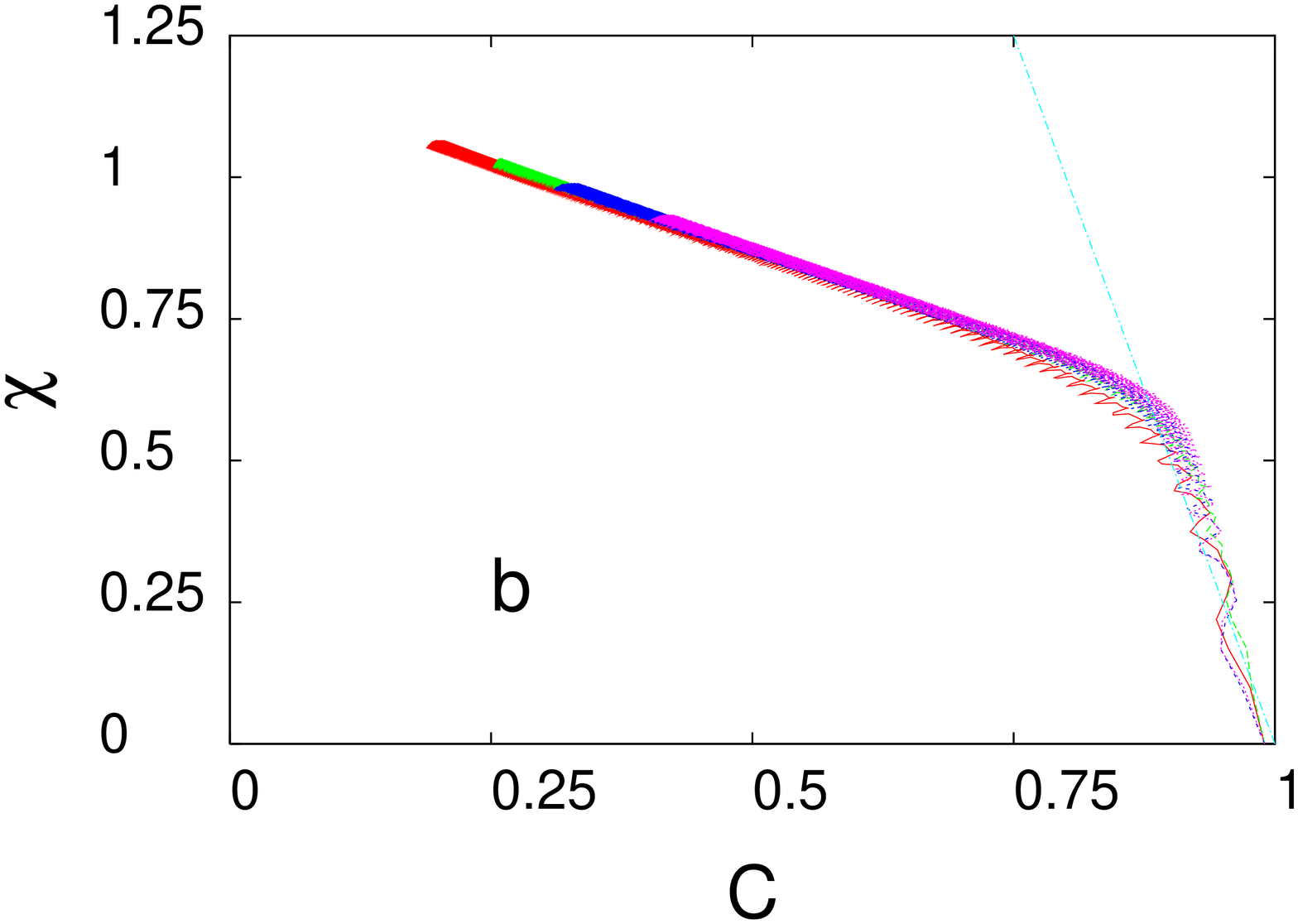,width=5cm,angle=-90}}
}
\caption{The $\chi(C)$ curves at $T=0.2$ with an applied ac-field 
of strength $h=1$ and frequencies $\omega=0.1$ (liquid phase) and 
$\omega=10$ (glassy phase), for different waiting times
$t_w=6,12,24,49$.}
\label{FDT_T02}
\vspace{.5cm}
\centerline{
\hbox{\epsfig{figure=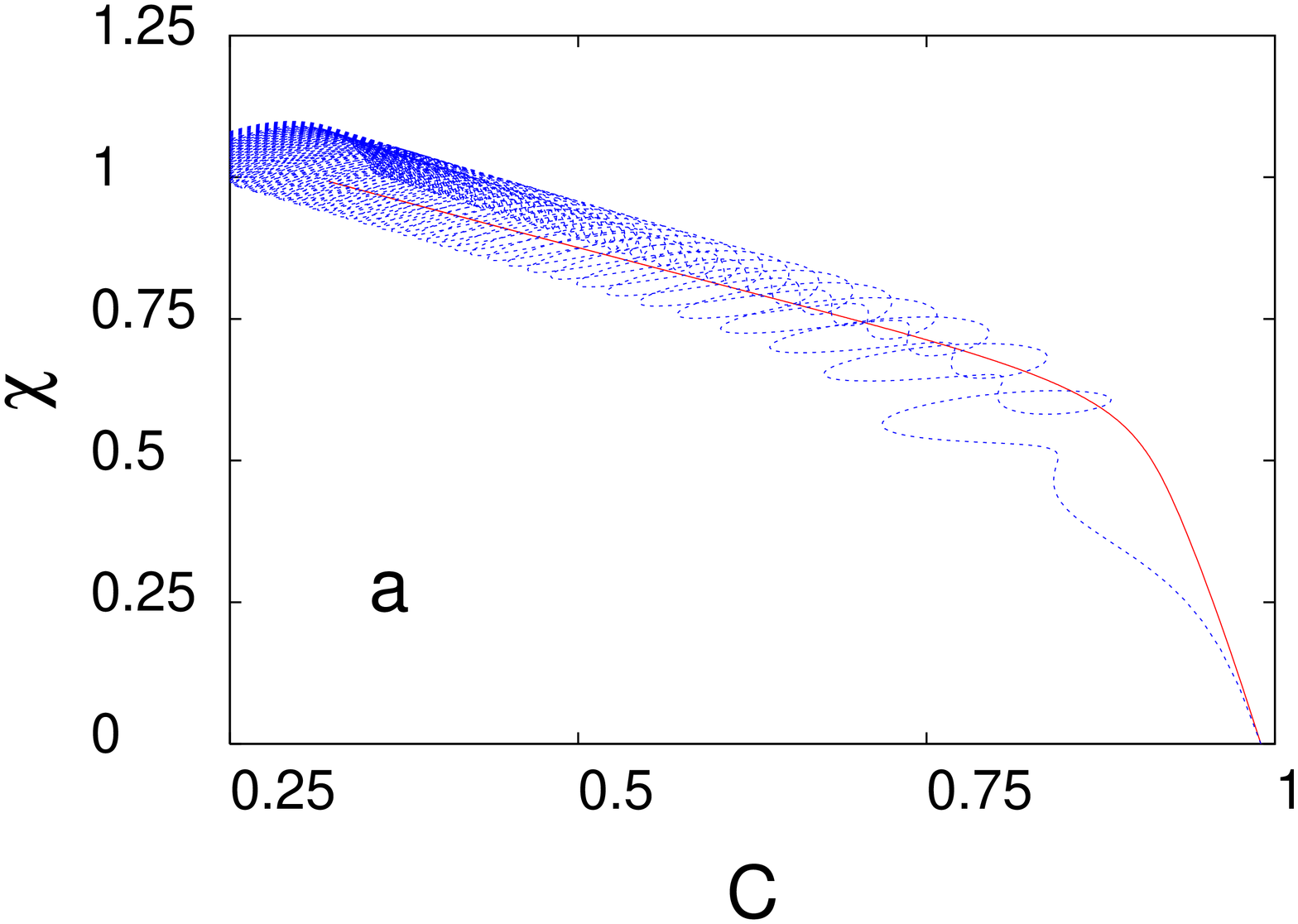,width=5cm,angle=-90}}
\hspace{.25cm}
\hbox{\epsfig{figure=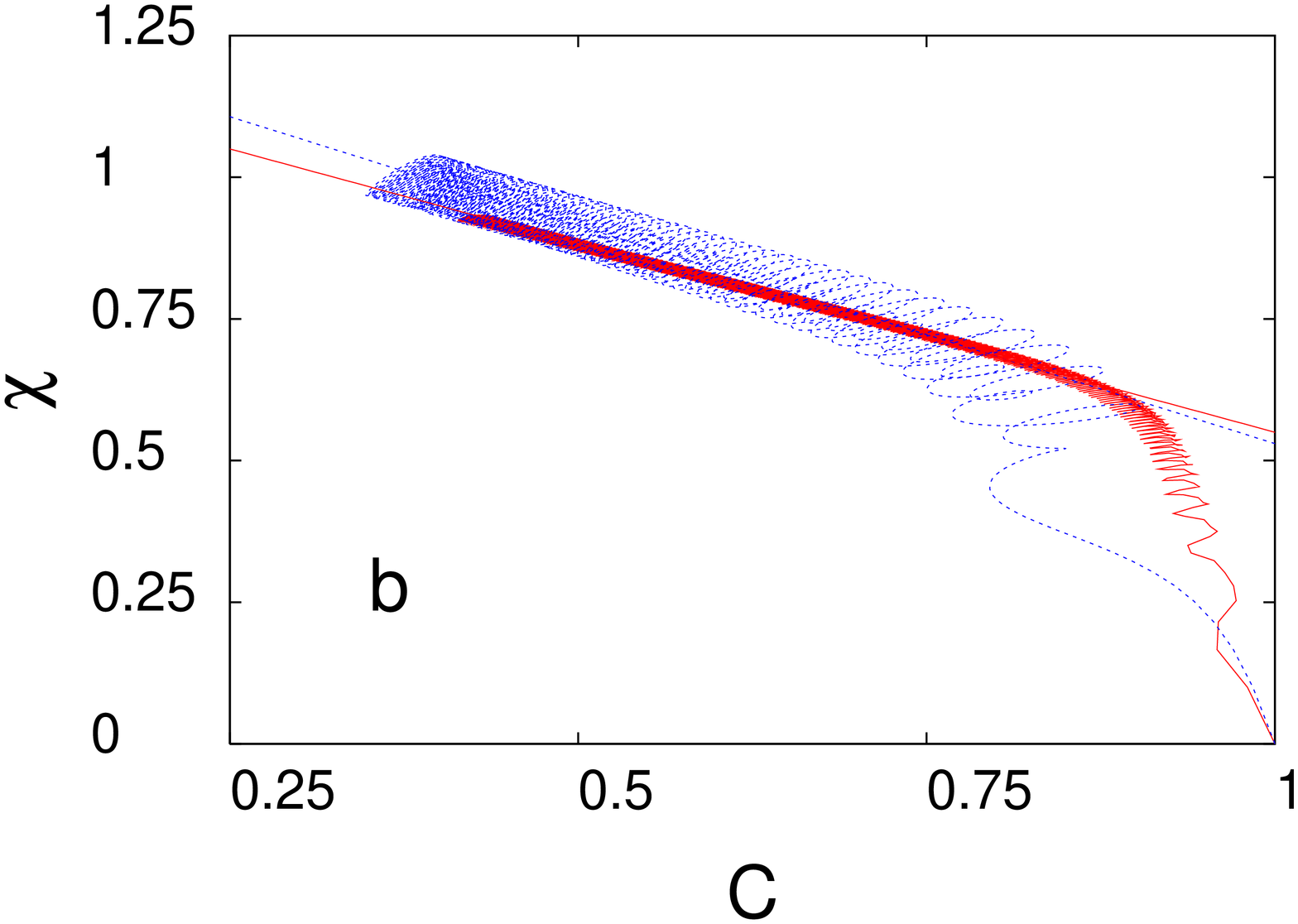,width=5cm,angle=-90}}
}
\caption{The $\chi(C)$ curves at $T=0.2$. Panel a, 
the two curves correspond to $h=0$ and $h=1$, $\omega=1$. In  both
cases the system is in its glassy phase.
In panel b  the two curves correspond to $\omega=1$ and $\omega=10$, 
both with an amplitude $h=1$, 
The lines are guides to the eye and they have the slope of the second 
part of the decay. $t_w=24$ for all curves displayed.}
\label{Teff_dep}
\end{figure}

We turn now to the study of the  {\sc fdt}. 
In Fig.~\ref{FDT_T02}
we show the evolution of the $\chi$ {\it vs} $C$ plots with $t_w$.
In Fig.~\ref{FDT_T02}-a, the system is not in the glassy
phase. 
Strong violations of the {\sc fdt} are observed, 
but no effective temperature can be defined. 
The system is athermal.

In Fig.~\ref{FDT_T02}-b, on the contrary, the system
is in its glassy phase.
One recovers then a $\chi(C)$ curve 
that is very similar to the one in zero field. 
The first part is almost a 
straight line with slope $-1/T$ while the second decay follows 
a temperature $-1/T_{\sc eff}$.
As  mentioned in Section~3, the fact that for small time scales, 
the {\sc fdt} is nearly satisfied results directly from 
the bound~(\ref{bound}).

The non-trivial outcome of this study  
is the fact that a well-defined $T_{\sc eff}$, which
exists in the non-driven system, may still be defined 
below the dynamical transition, in presence of the drive.
A stroboscopic construction, as the one in Section~3.3, will yield
a perfect straight line that defines $T_{\sc eff}$ unambiguously.

The actual value of the effective temperature may depend on 
various parameters.
We have searched for dependences on the frequency and 
amplitude of the ac field.
Within the range of parameters we could numerically
explore, we have not observed large dependences.
Figure~\ref{Teff_dep} display the $\chi$ {\it vs} $C$ 
curves for two choices of field amplitudes, and for two choices of
angular velocity at fixed field amplitude:
the dependence is indeed weak.

\subsection{Beyond mean-field: Effect of trapping states}
\label{beyond}

In this Section we present results from a numerical simulation
of the Ising version of the $p$-spin model in an ac field,
in the particular case $p=3$. 
We focus on the time dependence 
of the energy density $e(t) = N^{-1}\sum_{i<j<k} J_{ijk} s_i s_j s_k$
and the magnetization density 
$m(t) = N^{-1} \sum_i s_i(t)$.
The sizes $N=50$ and 150 have been used, together with 
a constant temperature
$T=0.01$, in all simulations. 
The amplitude and the frequency of the magnetic field have been varied.

The interest of such an investigation is that the finite $N$ behavior
of the model is accessible.
The system is hence able to escape and visit 
the trapping states described in Section~\ref{definition} 
in a finite times~\cite{Cukulepe}.
For that reason, the results presented in this section cannot be obtained
in a mean-field approach.

\begin{figure}[h]
\begin{center}
\begin{tabular}{cc}
\epsfig{figure=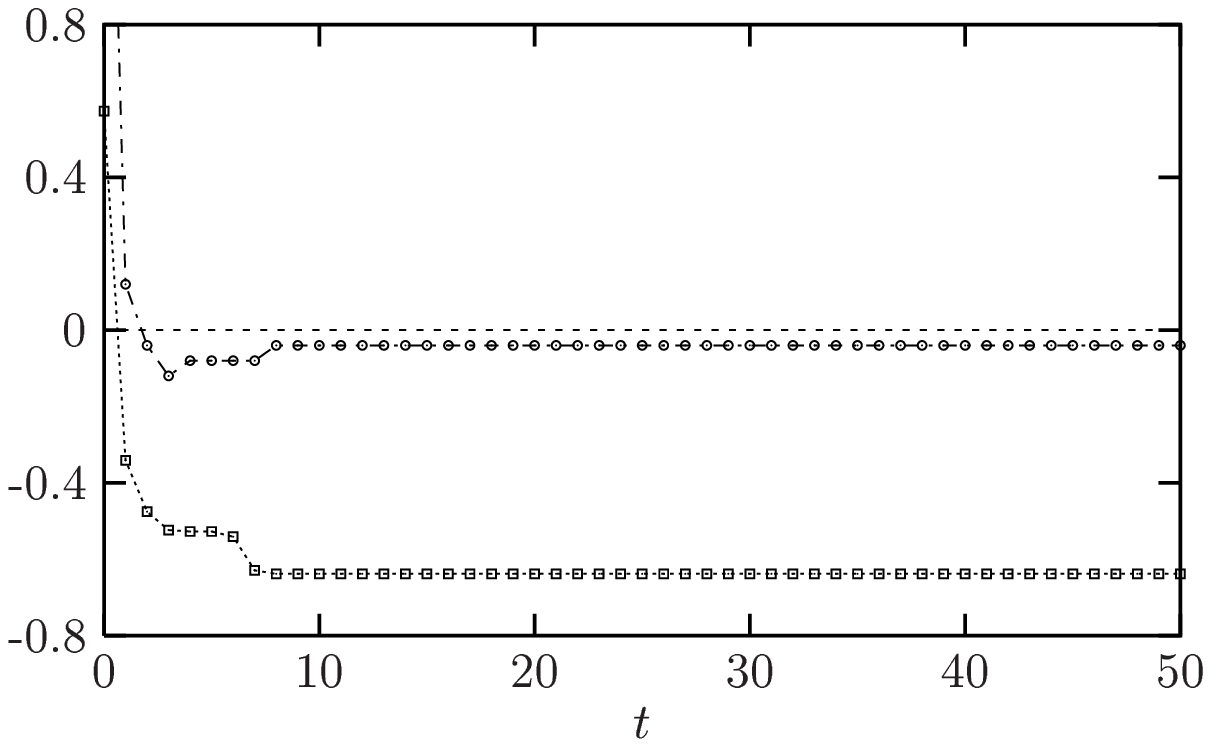,width=7.5cm,height=5.5cm} &
\epsfig{figure=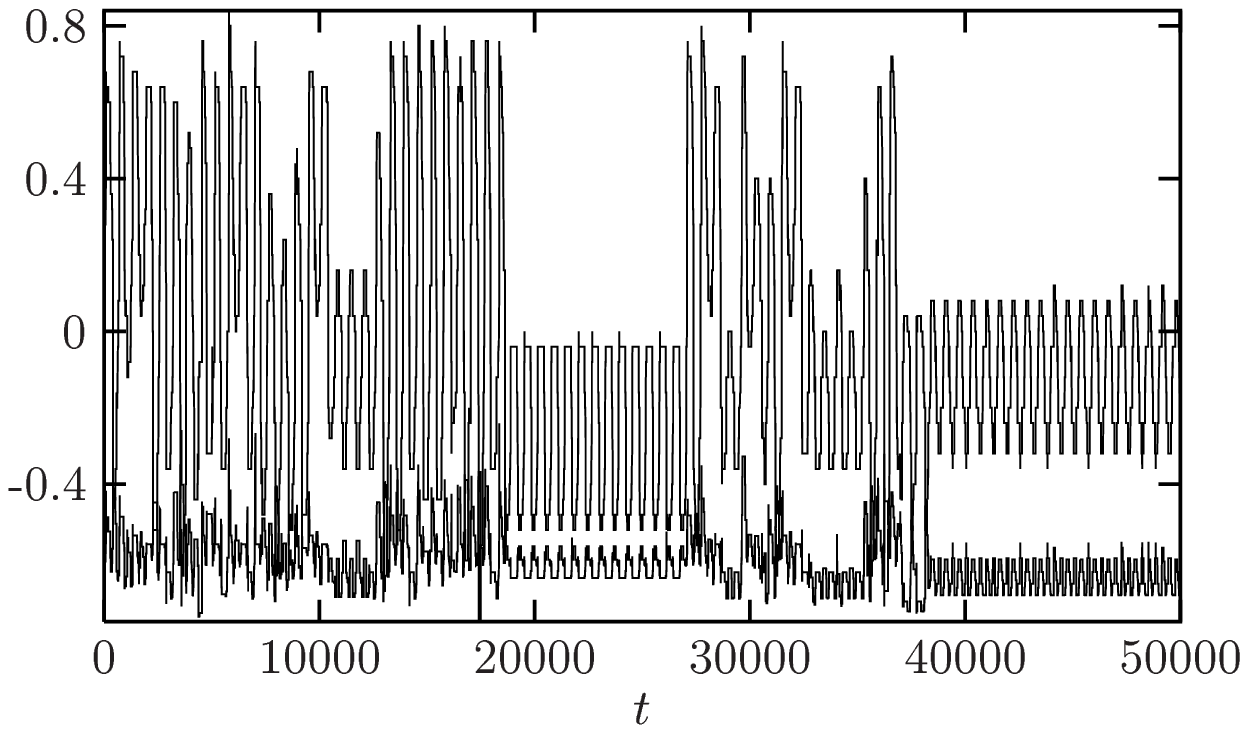,width=7.5cm,height=5.5cm}
\end{tabular}
\caption{Evolution of the energy (bottom curve) and magnetization (top curve)
densities without (left) and  with (right) magnetic field.
The parameters of the field are $h=2$, $\omega=0.01$.
Note that the time range in the left figure is much shorter, since both
$e(t)$ and $m(t)$ are constant for times $t>50$.}
\label{traps1}
\end{center}
\end{figure}

\begin{figure}
\vspace{.5cm}
\begin{center}
\begin{tabular}{cc}
\epsfig{figure=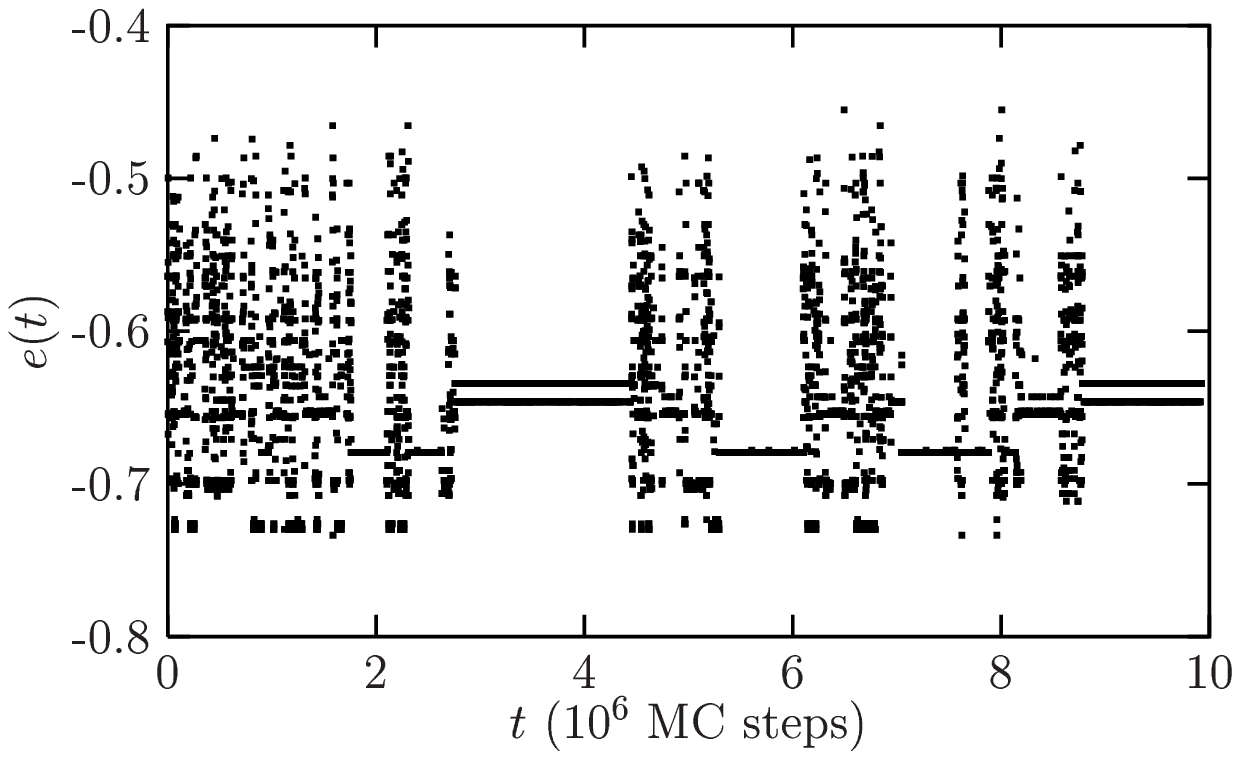,width=7.5cm,height=5.5cm} &
\epsfig{figure=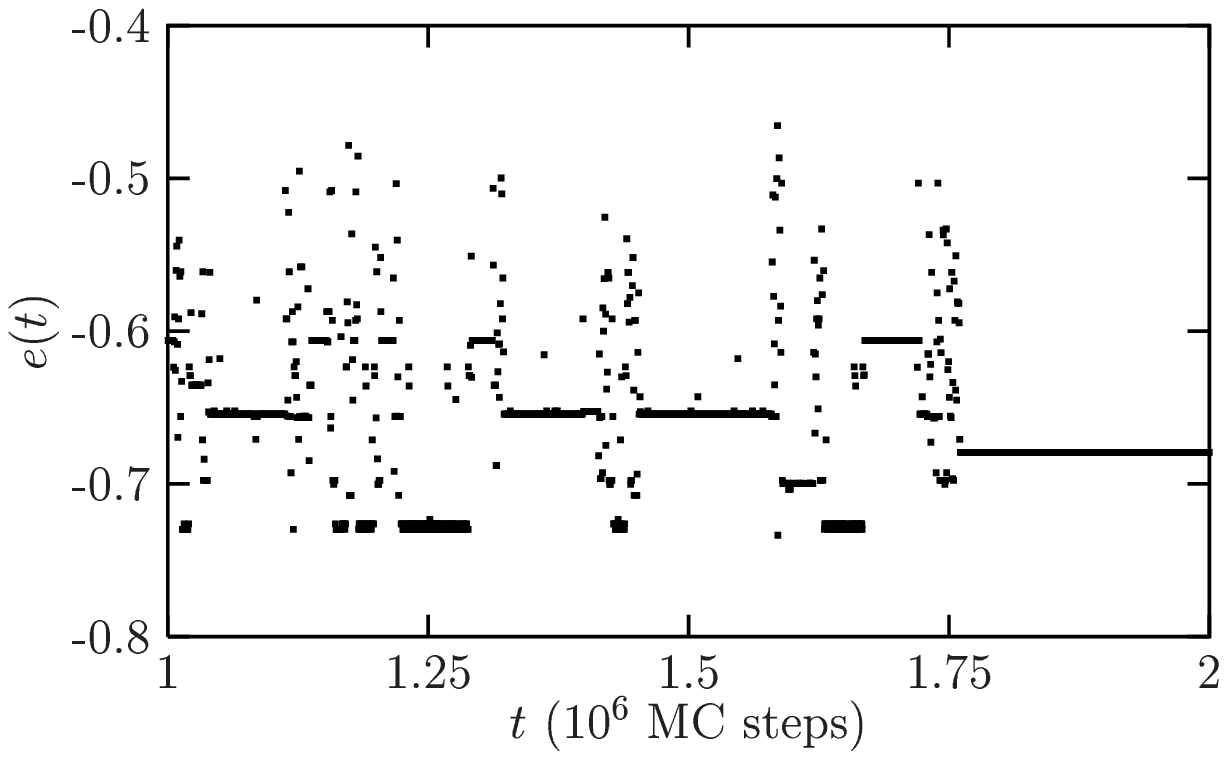,width=7.5cm,height=5.5cm} \\
\end{tabular}
\caption{Same parameters as in Fig.~\ref{traps1}. 
Evolution of the energy density with 
a magnetic field. Only one
point per period is represented.} 
\label{ene_traps2}
\end{center}
\end{figure}

In the absence of magnetic field, 
the energy density rapidly approaches
its asymptotic value.
At such a low temperature, the system
reaches a metastable stable, characterized by a nearly zero 
magnetization density, and cannot escape
within the numerical time window.
Fig.~\ref{traps1} (left) illustrates this `jammed' behavior.
The value of the energy density depends on the initial conditions,
which means that the system may be blocked in different 
trapping states when it starts from different initial conditions.

The influence of an ac magnetic field on this jammed behavior is
evident in Fig.~\ref{traps1}.
Both the energy and the magnetization density are 
oscillating functions, and their behavior is intermittent.
For some time windows, both quantities have an evolution which 
varies considerably from period to period.
For other time windows, on the contrary, the different 
cycles are very similar.
This can be interpreted as being due to  
the presence of the trapping (`jammed')
states: the system usually evolves inside one
of the trapping states, without escaping, and all the periods are
equivalent.
But from time to time, the driving force is able to make the system
escape the state, and the evolution is very erratic until
another trapping state is found.

This is confirmed in Fig.~\ref{ene_traps2}, where
the energy density is represented as a function
of $t$ for times of the form $t = n \tau$. 
The long horizontal plateaux are the moments where
the system is jammed, whereas
between the plateaux, the energy density changes values very rapidly. 
As in Ref.~\cite{Cukulepe}, we show in Fig.~\ref{ene_traps2}
that the time evolution of $e(t)$ is `self-similar',
in the sense that zooming on a time window makes smaller
plateaux become visible while the overall evolution looks the same.

This intermittent behavior is clearly reminiscent of the behavior
encountered in some granular experiments, where the
powder is very slowly perturbed, like the ones performed by
the Jussieu group~\cite{Kolb}.
In particular, it would be very interesting to perform
a more careful analysis of the statistical properties of 
$e(t)$ like, {\it e.g.}, measuring the statistics of 
trapping times~\cite{Kolb}.
We note also that this phenomenology is very similar to
the one of the so-called `trap model'~\cite{Bou}, that has been 
extended by Head to describe the phenomenology of granular
materials~\cite{head2}.

\begin{figure}[t]
\begin{center}
\begin{tabular}{cc}
\epsfig{figure=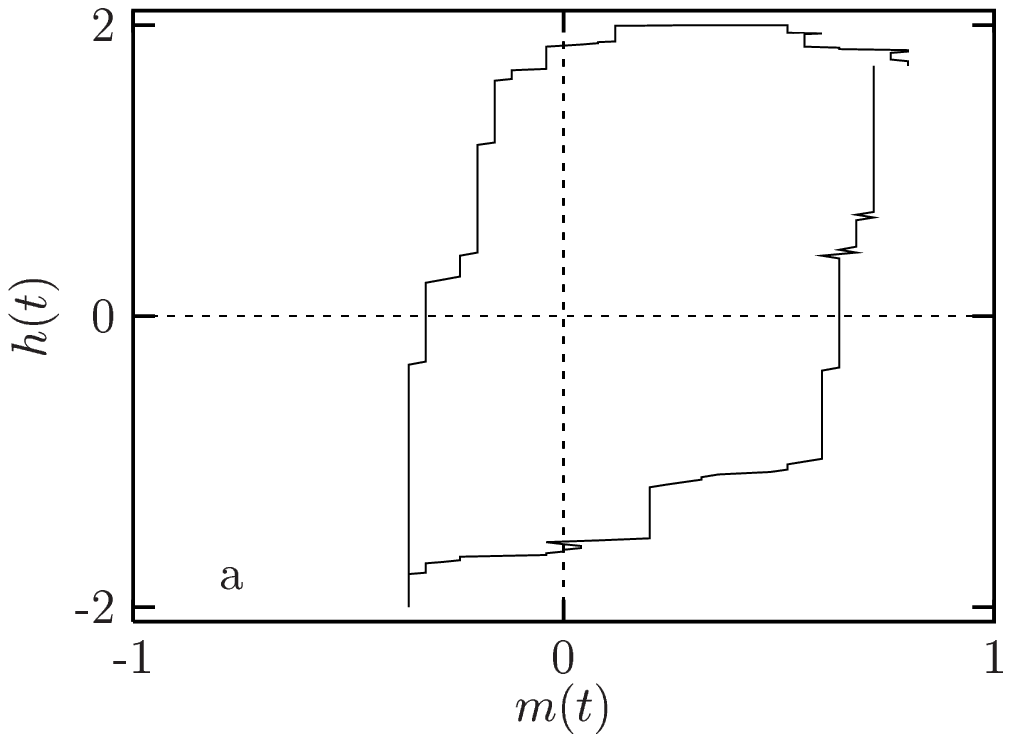,width=6cm,height=5cm} &
\epsfig{figure=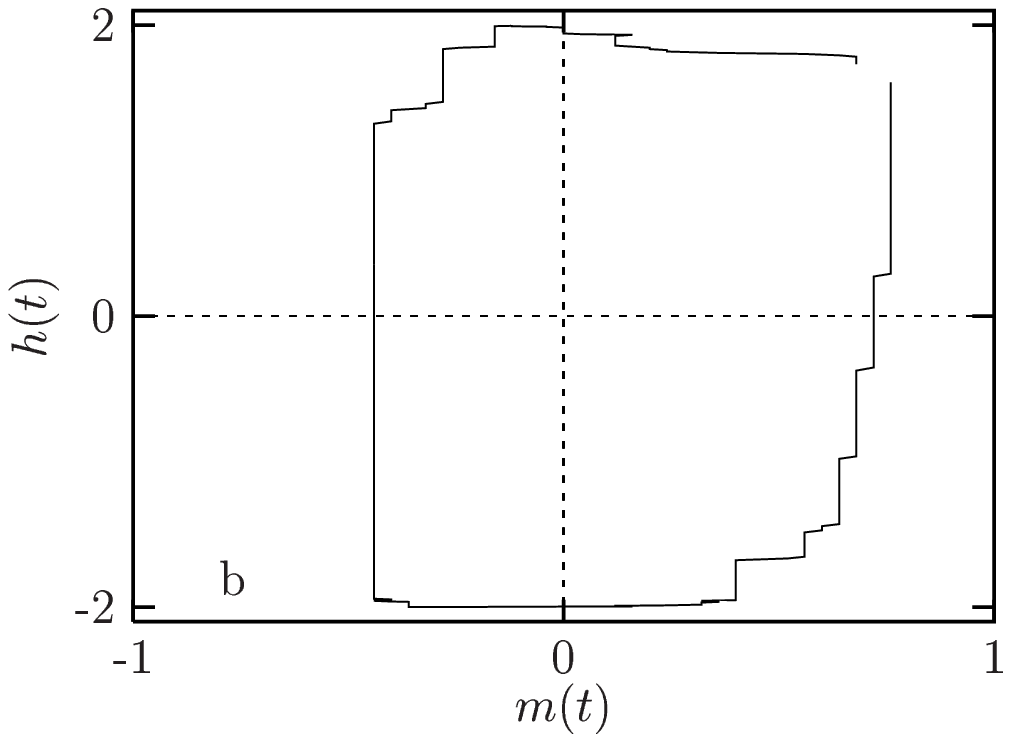,width=6cm,height=5cm} \\
\epsfig{figure=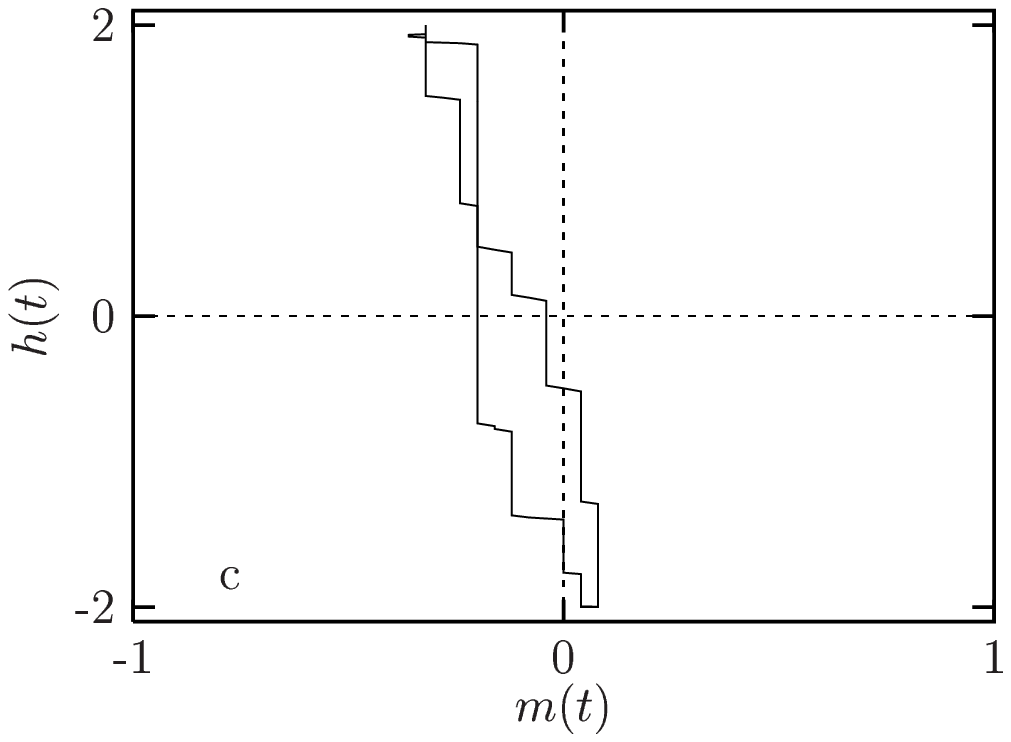,width=6cm,height=5cm} &
\epsfig{figure=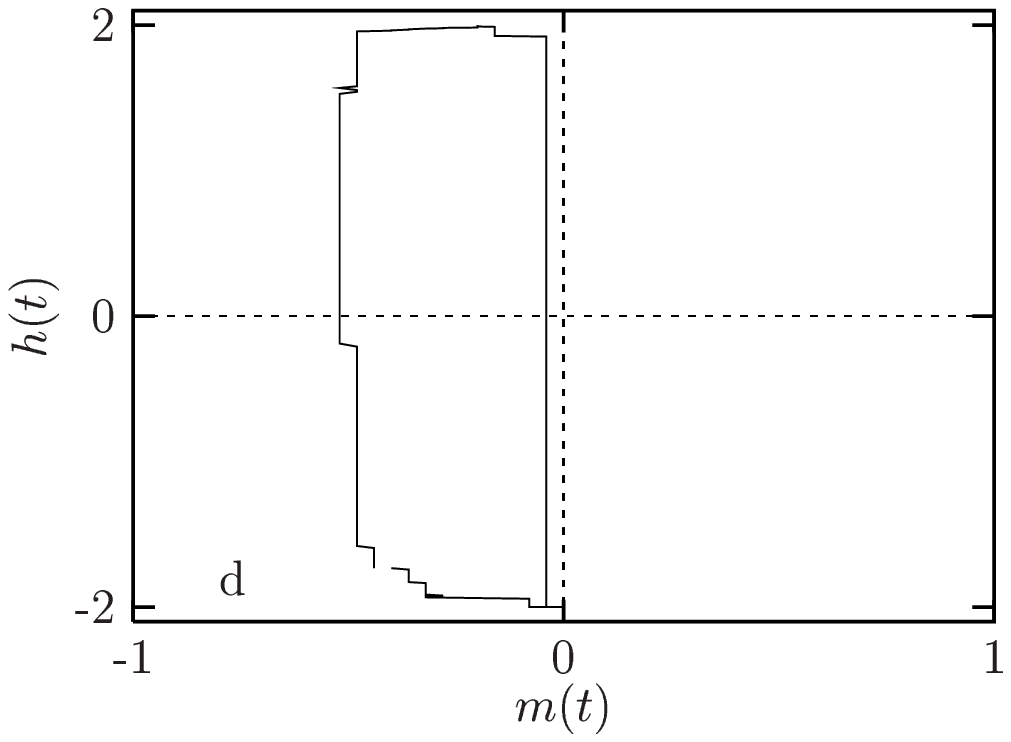,width=6cm,height=5cm} \\
\end{tabular}
\caption{Hysteresis loops. Panels a and b correspond to the period 
of rapid motion while panels c and d correspond to the interval 
when the system is trapped. $N=50$, $T=0.01$, $h=2$ and $\omega=0.1$}
\label{loops}
\end{center}
\end{figure}

Finally, it is interesting to stress that even within one cycle
of the field, the evolution is not regular at all.
This can be nicely seen in Fig.~\ref{loops}, where
the field $h(t)$ is represented as a function of the magnetization
$m(t)$.
This is the usual view of a hysteresis loop 
in ferromagnetic systems.
It is clear that the shape of the loops is far from elliptic, 
and that the system evolves in steps, rather than continuously:
this is analogous to the Barkhausen noise~\cite{sethna}.
The overall shape of a cycle drastically depends
on the fact that the system is trapped or not.
It is beyond the scope of this paper to study these loops
in detail, but we emphasize that this (mean-field) model
could be an interesting starting point to study 
the Barkhausen noise, in the spirit of Ref.~\cite{sethna}.

\section{Summary and link with granular matter}
\label{discussion}

\begin{figure}
\begin{center}
\psfig{file=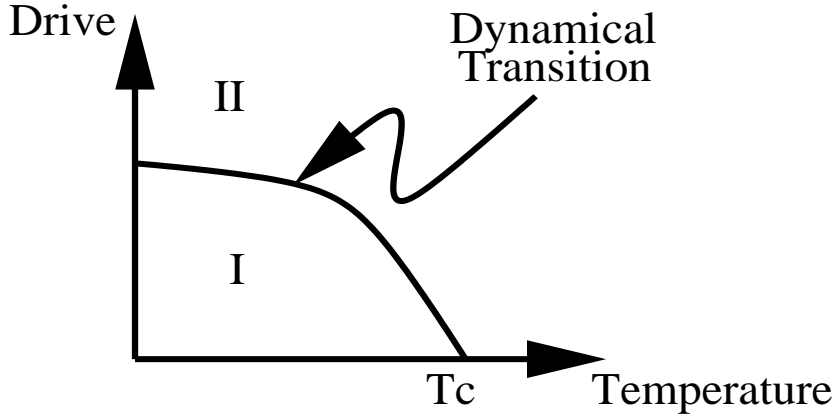,width=11cm,height=5.5cm}
\caption{Temperature-Drive phase diagram of a glassy system.
The vertical axis represents the typical amplitude of the driving force: its
precise definition depends on the specific experiment considered;
$T_c$ is the zero-drive glass transition temperature. }
\label{phasediagram}
\end{center}
\end{figure}

{\it Phase diagram. ---}
Our findings concerning the behavior of glassy systems 
under a time-dependent driving force are summarized in Figs.~\ref{sketch} and 
~\ref{phasediagram}. For any fixed frequency of the applied field, 
and a temperature $T<T_c$, where $T_c$ is the zero-drive glass 
transition temperature,  there are two regimes:
(I) At small drive, the system exhibits slow (``glassy'') dynamics.
(II) At large drive, there is no slow dynamics.
There exists then a well-defined critical drive separating these
two regimes.
On the other hand, if we work at fixed drive and modify the 
angular velocity, the system undergoes a transition from a 
liquid-like phase at low (though non vanishing) frequency to a 
glass-like phase at high frequency. This result is displayed 
in Fig.~\ref{sketch}.

The phase diagram in Fig.~\ref{phasediagram} is 
the analog to the $(T,\Gamma)$ plane 
of the three-dimensional phase diagram proposed 
by Liu and Nagel~\cite{liu_nagel}. 
We have then investigated its properties in detail, 
which had not been done so far in the case of 
a time-dependent driving force.
Contrary to previous phenomenological modeling of the dynamics of granular
matter assuming that the drive $\Gamma$ is related (in 
a possibly nonlinear way) to the temperature $T$ of a glassy 
model borrowed from statistical mechanics, 
we propose here to study precisely the interplay between
$T$ and $\Gamma$ as an intermediate step from glassy 
to granular materials.
The conclusions we draw are then directly pertinent
to gently driven granular materials or slow granular rheology.

We emphasize once again, that 
the model we have studied is {\it not} intended to describe granular matter 
in full microscopic detail, but rather 
to act as a source of inspiration for the interpretation of 
numerical and experimental results.
It may also motivate new  experimental measurements. 
Let us however briefly summarize its dynamic behavior 
in the context of existing data for granular matter.
 
{\it One-time quantities. ---}
In phase (I), one-time quantities (we have focused
on the energy density) typically decay towards
their asymptotic values as power laws.

It is experimentally established that 
the density $\rho(t)$ in a gently driven granular system  exhibits a 
very slow relaxation~\cite{reviews}.
The parallel can be drawn with the energy density $e(t)$ (or
more precisely with $-1/e(t)$) in glassy  models. 
The seminal experiments at Chicago have exhibited
a logarithmic relaxation of the density~\cite{compaction_exp}, as found in 
all subsequent 
works~\cite{jaeger,tetris0,tetris1,alain,mauro,boutreux,parking,linz,gavrilov,head1,head2}.
(Note that Ref.~\cite{compaction_exp} explicitly 
excludes a power law behavior.)
However, the recent experiment of 
Nicolas {\it et al.}~\cite{pouliquen} showed that 
the logarithmic law is {\it not} a universal behavior and that the 
time dependence may depend on the way the system is perturbed.
From the theoretical point of view, one could prefer a power law 
decay for any one-time quantity 
since it does not involve an intrinsic time scale, 
whereas the logarithmic law $1/\ln (t/t_0)$ explicitly
introduces a time scale $t_0$.
In this context, it is interesting to note that Barrat and 
Loreto~\cite{alain}, in their study of the Tetris model, 
have shown that the fitting parameter $t_0$ actually 
depends 
on the considered time window; qualitatively $t_0 \propto t_w$. 
Head~\cite{head2} had already emphasized the role of the time window 
on the fitting parameters,
in particular on the asymptotic value of the density.
 
In the context of glassy systems, it is a well-known feature that
the interpretation of data over a finite time-window 
is always delicate.
A typical example is provided by the case of dipolar 
glasses where 
the dielectric constant for samples prepared with 
different cooling rates seems to approach different asymptotic 
values~\cite{Levelut}.

{\it Two-time quantities. ---}
In phase (I), the model we have studied also exhibits aging.
For $p=2$, we find a  $t/t_w$-scaling (in stroboscopic time)
of the autocorrelation function while for $p\geq 3$ a much more careful 
numerical analysis is needed to determine this law.
Other glassy models perturbed with ac-forces may lead 
to other aging scalings.
Generally speaking, theoretical arguments~\cite{review_aging} 
indicate that one can expect the following behavior
for any two-time function $F(t,t_w)$
\begin{equation}
F(t,t_w) \sim {\cal F} \left( \frac{h(t)}{h(t_w)} \right),
\label{scaling}
\end{equation}
in a given time-scale. 
${\cal F}$ and $h$ are two model/system-dependent scaling functions.
More complicated scalings, such as a sum of different terms like
Eq.~(\ref{scaling}) implying several different time scales
in the problem, are also possible.

Two-time quantities such as the density-density
correlation function have been  numerically studied, clearly demonstrating that
{\sc tti} is lost~\cite{tetris1,parking,alain}:
the Tetris model exhibits
a $\ln(t)/\ln(t_w)$ behavior~\cite{tetris1,alain},
whereas the parking lot model shows a $t/t_w$ aging behavior~\cite{parking}.

An experimental determination of the scaling of two-time quantities will
be useful to discriminate among the different models
proposed to describe granular matter. 
This has already been proposed in Refs.~\cite{tetris1,alain,parking} and
amounts to an experimental determination of the function 
$h(t)$ involved in Eq.~(\ref{scaling}).
Such experimental determinations in glassy systems
are by now numerous.
They have emphasized on the one hand
the universality of the aging phenomena in many microscopically
different systems~\cite{Struick,spin-glasses,Leheny,Levelut,gel}, 
but on the other hand they have also revealed
subtle differences between them~\cite{review_aging,spin-glasses}.

{\it Fluctuation-dissipation theorem violations. ---}
Our study has shown that modifications of  the {\sc fdt}, rather 
similar to those observed for {\it aging} glassy systems, also arise
in glassy models driven by time-dependent forces.
The parallel with granular materials suggests
that this may also happen in driven powders.

The parametric plot we have built clearly shows that
driven glassy systems have two distinct time scales, 
which we interpret as in Refs.~\cite{mehta,mehta-barker}.
A fast one, which is essentially independent of the waiting time,
represents the individual motion of grains: no effective temperature,
in the sense of Eq.~({\ref{XFDT}), can be here defined.
Since the model we have used is itself thermal (it uses a Langevin equation
with a thermal noise) the fast motion has a reminiscence of the 
temperature of the bath (that may be zero). In a more realistic 
model of granular matter, it is likely that the fast motions
will be completely athermal.       
On the contrary, the long time scale growing with the waiting time
and giving rise to the  $t/t_w$-relaxation,
represents large structural rearrangements in the powder.
Our results indicate that these slow degrees of freedom are thermalized
at a well-defined effective temperature.  
We are confident that this 
result also holds in realistic models of granular matter.
 
Again, numerical and experimental work could be used to check
these predictions in granular materials.
We have shown that the two models studied here 
have different {\sc fdt} violations:
the $p=2$ case has an infinite effective temperature, whereas
the $p \ge 3$ models have a single finite effective temperature.
Other glassy models (for instance mean-field spin glasses) have
been shown to have a hierarchy
of effective temperatures when they are undriven, and shall
probably keep this behavior under driving forces~\cite{BBK2}.
We conjecture that granular materials fall in the $p \ge 3$ category,
by analogy with structural glasses. This prediction
has to be experimentally and numerically tested.
It is not unlikely that different microscopic models lead to
different {\sc fdt}-violations; such a numerical determination
could once again discriminate between them.

To study {\sc FDT} violations during the compaction experiment,
one has to compute, separately, the correlation function between two
observables and its conjugated response function.
In experiments, for instance, the Nyquist relation is checked
through dielectric measurements~\cite{Leheny,ludo}.
In granular materials, a natural choice is to check 
the Einstein relation between self-diffusion and mobility of the grains.
The results obtained in Refs.~\cite{mauro2,makse} are then very
encouraging, although the driving field is not time-dependent in these studies.
Nicodemi studied instead the fluctuation-dissipation
relation between height-height displacement and its conjugated 
integrated response to small shaking amplitude perturbations in the 
Tetris model~\cite{nicodemi}. 
In the low density regime (fluide-like), after a short transient,
these two quantities are related by the usual FDT. In the high
density regime, the relation between these quantities 
shows much stronger deviations from FDT 
than the model we studied here. 
This difference may be due to the kind
of perturbation used by Nicodemi. 
For a thermal system, the procedure in Ref.~\cite{nicodemi}
is indeed similar to measuring
the energy response to a temperature change: The resulting
{\sc FDT} properties are found to be of an unusual
type~\cite{suzanne} also in this case. However, the notion of 
an effective temperature~\cite{Cukupe} implies that the FDT relations 
of all observables evolving in the same time-scale should be identical. 
As suggested in Refs.~\cite{alain,mauro2}, an interesting open 
question is:
does a measurement in the Tetris model done with a more standard 
infinitesimal perturbation lead to the same {\sc FDT} relation? This point 
deserves a more detailed investigation in this and other glassy models. 

{\it The dynamic transition, and the regime} (II). ---
We have found a well-defined transition in the phase diagram
of Fig.~\ref{phasediagram}.
This is in complete agreement with experimental and
numerical observations
of  the existence of a critical value of the drive $\Gamma^\star$,
below which glassy effects may be observed.
The existence of this dynamical transition
provides a very natural context for the 
interpretation of the numerous non-equilibrium
effects encountered in granular matter.
Indeed, the transition line in the Fig.~\ref{phasediagram} links 
the standard glass transition arising at $T=T_c$, $\Gamma=0$ in the glassy 
model to the dynamical transition arising at $T=0$, $\Gamma=\Gamma^\star$
in the driven dynamics of athermal systems.
In our opinion, this result gives a nice theoretical support to 
the glasses/granular analogy, making a deep connection 
between the two situations.

We wish to emphasize that
it was not evident {\it a priori} that the transition 
could survive a finite time-dependent driving force. This is
one of the main results of this paper and 
it has to be confronted to the very different effect of shear-like forces. 
Indeed, an infinitesimal shear-like perturbation,
is enough to introduce a finite time-scale in this model,
and aging is hence interrupted~\cite{Cukulepe,BBK}.  

The liquid regime (II) is less interesting in the present context, 
since no slow dynamics is present.
Let us  stress that the system is
still strongly driven, and hence no `equilibrium' state,
in the sense of statistical mechanics, is reached.
In particular, no effective {\sc fdt} temperature can be defined:
the system is completely athermal.
The very existence of an effective temperature relies indeed 
on the existence of two well-separated time scales allowing 
the slow degrees of freedom to thermalize and the 
susceptibility / correlation parametric  curves of Fig.~\ref{FDT_T02}-a
are not very useful to describe this non-equilibrium situation.

\section{Conclusion: towards a thermodynamical 
description of the slow granular rheology?}
\label{conclusion}

The interpretation of {\sc fdt} violations in terms of
an effective temperature $T_{\sc{eff}}$ 
has been developed in Ref.~\cite{Cukupe}.
Having a well-defined concept of temperature is a crucial first step
for a thermodynamical description of granular materials~\cite{Theo_silvio}.
It is then important to discuss the possibility
of a link with previous thermodynamical concepts in the 
granular literature. 

The notion of a `granular temperature' has been introduced
to extend thermodynamics to strongly perturbed powders~\cite{reviews}. 
In analogy with the kinetic gas theory, it has been
assumed that distribution of the grain velocity $v$ is Maxwellian
and via the equipartition relation $T \propto \langle v^2\rangle$.
We have been concerned with the weakly perturbed 
regime, where the dynamics is slow. 
This hydrodynamic definition is not supposed to 
be relevant in this case. 
This is also well known in the field of glasses
where nontrivial effective temperatures are known to exist while the 
kinetic energy is an observable that very rapidly equilibrates with its 
environment, leading to a kinetic temperature
that coincides with the temperature  of the bath.
Very recently~\cite{lyderic}, an attempt to extend the hydrodynamic theory
to a regime where the grains are {\it not} fluidized has been
proposed, also using the concept of kinetic granular temperature.
It would then be very interesting to try to understand the two concepts
in a unified way.

More related to ours is the approach of Mehta {\it et al.}~\cite{mehta},
discussed in the introduction.
This phenomenological two-step, two-temperature model finds a nice
justification within 
the scenario emerging from our results.
This model was in fact inspired by the 
illuminating work initiated in the late 80's by 
Edwards and coworkers~\cite{edwards}.
They have postulated thermodynamic relations in analogy with
the usual `thermal' thermodynamics, where the 
volume $V$ plays the role of the internal energy $U$.
In this approach, the entropy is the logarithm of 
the number of configurations with volume $V$.
The so-called compactivity, $X$, is the analog of the temperature
and it then defined as~\cite{edwards}
\begin{equation}
\frac{1}{X} \equiv \frac{\partial S}{\partial V} \; , 
\label{Xedw}
\end{equation}  
still by analogy with the usual definition
$T^{-1} \equiv \partial S/\partial U$.
The connection between the compactivity $X$ and 
the {\sc fdt} temperature $T_{\sc{eff}}$
discussed in this paper has already 
been explored~\cite{jorge,Baetal,nicodemi2,Monasson,nicodemi}. 
Crucially for our study, {\it Edwards' definition of the compactivity 
coincides with the (asymptotic) {\sc fdt} 
effective temperature} defined in Eq.~(\ref{Teff}).
This result holds
for mean-field models like the one discussed here~\cite{Monasson},
as well as in the lattice gas model
studied in Ref.~\cite{Baetal}. However, these studies
have been done without an external forcing. 
It would be very interesting to extend the numerical 
analysis of Ref.~\cite{Baetal} to the 
same model now driven by an ac force. 

We have obtained here that the existence of an effective temperature 
for the slow decay resists
a finite ac force. This supports the conjecture
that the definitions (\ref{Teff}) and (\ref{Xedw}) may 
be of fundamental interest for the study 
of glassy/granular materials. 
Computing them in realistic models is then a challenge
for future research~\cite{Baetal,nicodemi2,Monasson,jorge_giulio}.
In this respect, very recent works~\cite{anita} 
studying spin models on random graphs
may give some insights on the role played by metastable 
states, which can have an important 
influence on the dynamics, as we showed in Section~\ref{beyond}

\vspace*{0.3cm} 

In conclusion, we have found that the study of driven glassy
systems provides a theoretical framework to understand the slow
granular rheology. The existence of a dynamical transition
justifies well the use of `modified glassy models' to 
describe granular materials.
The existence of an effective temperature for the slow degrees
of freedom provides 
in particular a nice theoretical basis to 
previous `thermal' models for granular matter,
and to the seminal approach of Edwards.

\section*{Acknowledgments}
We wish to thank J.-L. Barrat, E. Kolb, J. Kurchan, F. Restagno, 
M. Sellitto, L. Vanel and J. Wittmer for very useful discussions.

\end{document}